\documentstyle[twocolumn,psfig,aps]{revtex}
\begin{document}
\twocolumn[\hsize\textwidth\columnwidth\hsize\csname@twocolumnfalse\endcsname
\title{EXPERIMENTS ON LADDERS REVEAL A COMPLEX INTERPLAY BETWEEN A
SPIN-GAPPED NORMAL STATE AND SUPERCONDUCTIVITY}
\author{Elbio DAGOTTO}
\address{National High Magnetic Field Lab and Department of Physics,
Florida State University, Tallahassee, FL 32306, USA} 
\address{By invitation for Reports of Progress in Physics}
\maketitle

\begin{abstract}

In recent years, the study of ladder materials has developed into a
well-established area of research within the general context of Strongly
Correlated Electrons. This effort has been triggered by an unusual
cross-fertilization between theory and experiments. In this paper, the
main experimental results obtained in the context of ladders are
reviewed from the perspective of a theorist.  Emphasis is given to the
many similarities between the two-dimensional high-$\rm T_c$ cuprates
and the two-leg ladder compounds, including
Sr$_{14-x}$Ca$_x$Cu$_{24}$O$_{41}$ (14-24-41) which has a
superconducting phase at high pressure and a small hole
density. Examples of these similarities include regimes of linear
resistivity vs temperature in metallic ladders and a normal state with
spin-gap or pseudogap characteristics.  Some controversial results in
this context are also discussed.  It is remarked that the ladder
14-24-41 is the first superconducting copper-oxide material with a
non-square-lattice layered arrangement, and certainly much can be
learned from a careful analysis of this compound.  A short summary of
the main theoretical developments in this field is also included, as
well as a brief description of the properties of non-copper-oxide
ladders.  Suggestions by the author on possible experiments are
described in the text.  Overall, it is concluded that the enormous
experimental effort carried out on ladders has already unveiled quite
challenging and interesting physics that adds to the rich behavior of
electrons in transition-metal-oxides, and in addition contributes to the
understanding of the two-dimensional cuprates. However, still
considerable work needs to be carried out to fully understand the
interplay between charge and spin degrees of freedom in these materials.

\end{abstract}

\section*{CONTENTS}
\begin{tabular}{ll}
{\bf 1. Introduction}............................................................................................................................................ & 1 \\
{\bf 2. Theoretical Aspects: a Brief Summary}............................................................................................. &  3\\
\hspace{.6cm} 2.1 Undoped Spin Ladder Models............................................................................................................ &  3 \\
\hspace{.6cm} 2.2 Hole-doped Spin Ladder Models........................................................................................................ &  4 \\
{\bf 3. Experimental Results}............................................................................................................................   & 5\\
\hspace{.6cm} 3.1 The Cu-oxide Ladder Compound SrCu$_2$O$_3$....................................................................................... & 5\\  
\hspace{1.2cm} 3.1.1 Doping of SrCu$_2$O$_3$ with Zinc................................................................................................... & 6\\
\hspace{.6cm} 3.2 The metallic ladder compound
La$_{1-x}$Sr$_x$CuO$_{2.5}$............................................................................... & 7 \\
\hspace{.6cm} 3.3 The Superconducting Ladder Compound
Sr$_{14-x}$Ca$_{x}$Cu$_{24}$O$_{41}$........................................................... & 8\\
\hspace{1.2cm} 3.3.1 Superconductivity in Doped 14-24-41 with x=13.6.................................................................. &
8 \\
\hspace{1.2cm} 3.3.2 Superconductivity in Doped 14-24-41 with x=11.5.................................................................. &
9 \\
\hspace{1.2cm} 3.3.3 Optical Conductivity $\sigma(\omega)$........................................................................................................ & 11\\ 
\hspace{1.2cm} 3.3.4 Nuclear Magnetic Resonance.................................................................................................... & 12\\ 
\hspace{1.2cm} 3.3.5 Inelastic Neutron Scattering..................................................................................................... & 13\\
\hspace{1.2cm} 3.3.6 Photoemission and Angle-Resolved Photoemission ${\rm
A({\bf p}, \omega)}$..................................................... & 13 \\
\hspace{.6cm} 3.4 The Strong-Coupling Ladder Compound
Cu$_2$(C$_5$H$_{12}$N$_2$)$_2$Cl$_4$........................................................... & 14 \\  
\hspace{.6cm} 3.5 Vanadium-based Ladder Compound CaV$_2$O$_5$................................................................................... & 14 \\  
\hspace{.6cm} 3.6 The Ladder Compound KCuCl$_3$....................................................................................................... & 15 \\  
{\bf 4. Conclusions}............................................................................................................................................ & 15 \\
{\bf 5. Acknowledgements}............................................................................................................................... & 16 \\
\end{tabular}

\vskip2pc]
\narrowtext

\section{Introduction}

The theoretical and experimental study of materials with an ionic
structure containing dominant patterns in the shape of ``ladders'' has
recently attracted a considerable attention in the context of Strongly
Correlated Electrons and Condensed Matter physics.  The $n$-leg ladders
are defined as $n$ parallel chains of ions, with bonds among them such
that the interchain coupling is comparable in strength to the couplings
along the chains. The particular case of $n$=$2$ motivates the use of
the name ``ladder'' for this geometry. The coupling between the two
chains that participate in this structure is through ``rungs'', language
also extended to the $n$-leg systems.  A vast literature on this subject
has already accumulated, especially in the last three years. This made
the study of ladders one of the two ``hottest'' topics of research in
Condensed Matter for 1996 (see Greven, Birgeneau and Wiese, 1997; Levy,
1996) and investigations in this area continue at a rapid pace.

What is the cause for this sudden interest on ladder systems?  There are
basically two broad reasons that have triggered the present enthusiasm
on the subject. First, ladders provide a ``playground'' for studies of
high critical-temperature ($\rm T_c$) superconductors (Bednorz and
M\"uller, 1986) since in the absence of hole carriers they have a
$spin$-$gap$ in the energy spectrum, namely it costs a finite amount of
energy to create spin excitations above the ground state, the latter
being a spin-singlet. This property resembles the spin-gap feature that
has been observed in the high-$\rm T_c$ cuprates, particularly in the
underdoped regime of low hole-density. Actually, for the two-dimensional
(2D) cuprates a ``pseudogap'' is a better term for this feature since
low-energy spin excitations exist, although with low spectral weight in
neutron scattering experiments.  Since at hole concentrations below the
optimal value the normal state above $\rm T_c$ presents such a
pseudogap, it is expected that its existence should be important for the
superconductivity that occurs once the temperature is reduced below $\rm
T_c$. However, no clear consensus has been reached on the origin of the
pseudogap in the 2D cuprates.  In this framework, the ladder systems
provide an interesting simpler setup for the analysis of spin and charge
excitations in a spin-gapped environment since these systems certainly
have a gap, both in theoretical calculations as well as real
experiments, as discussed below.

Adding to this interesting spin-gap property, the early theoretical
studies of ladders showed that upon doping of holes the ground state
becomes dominated by $superconducting$ correlations, as explained in the
next section. Then, the analogy with the underdoped high-$\rm T_c$
cuprates is even stronger, since the ladders not only share with them
the presence of a spin-gap but also superconductivity is expected at a
low-concentration of holes. Moreover, theory predicts that this
superconductivity for ladders should be in the d-wave channel, the
currently most accepted channel for superconductivity in the high-$\rm
T_c$ cuprates, adding further evidence for strong similarities between
doped-ladders and doped-planes.  Last but not least, ladders systems are
considerably easier to study theoretically than two-dimensional models
because they are basically quasi-one-dimensional. A plethora of powerful
many-body techniques, notably those involving computational methods,
work well in one-dimension but loose their accuracy in two-dimensions
(Dagotto, 1994).  Currently it is believed that the physics of isolated
two-leg ladders is under reasonable theoretical control, and there is
little controversy on its main properties.

The second motivation for the sudden interest in ladder systems is
related with the explicit realization of ladder compounds reported in
seminal experimental papers reviewed here.  The motivation provided by
the theorists triggered an enormous effort on the experimental front to
synthesize ladder materials, searching for the two main predictions made
before those experiments, namely the existence of a the spin-gap and
superconductivity. As shown below and after a considerable effort, clear
evidence has already accumulated that real ladder materials with an even
number of legs, two in particular, have a finite spin-gap in their
spectrum of spin excitations in agreement with the theory. In addition,
superconductivity in one of the ladder compounds has been detected upon
the introduction of hole carriers and using high pressure. It is
certainly tempting to associate this superconducting phase once again
with the early predictions made by theorists in this context.  However,
it is fair to say that the origin of superconductivity in ladders is
still actively under investigation and a consensus on this subject has
still not been reached.  Of particular current interest is the possible
relevance of the interladder coupling to stabilize the superconducting
phase. Thus, it remains to be investigated whether this phase
corresponds or not to the theoretical predicted superconductivity based
on the analysis of isolated ladders.  In addition, several recent
experiments have revealed additional close analogies between
superconducting ladder compounds and high-$\rm T_c$ superconductors.
Adding to the spin-gap and superconducting properties, it has been
observed that there are regions of parameter space where the resistivity
of ladders is $linear$ with temperature, a hallmark of the exotic normal
state found in the two-dimensional cuprates.

The large number of experiments already carried out in the context of
ladder materials and their many common features has motivated the
presentation of this review. The paper is focussed on the $experiments$,
aside from a brief summary of the main theoretical results in the next
section, and it certainly does $not$ attempt to review the huge
literature accumulated in recent years on theoretical studies of ladder
compounds. Such a formidable task is postponed for a future, more
comprehensive, review article. The organization of the present paper is
in sections that individually summarize the main results corresponding
to particular ladder materials. Particular emphasis is given to
copper-oxide based ladders, since some of them accept hole doping and
the physics of carriers in spin-gapped backgrounds can be
studied. However, other compounds are also described in the present
paper. Comments by the author on some experiments are included in the
text, and a comprehensive summary is provided in the last section of the
review. The overall conclusion is that the study of ladders has already
provided several interesting results to the experts in Strongly
Correlated Electrons, and further work in this context should certainly
be encouraged, especially regarding the clarification of the properties
of its superconducting phase.

\section{Theoretical Aspects: a Brief Summary}

\centerline{\bf 2.1 Undoped Spin Ladder Models}
\vspace{.6cm}

After a plethora of theoretical studies of the S=1/2 Heisenberg model
with interactions between nearest-neighbor spins, it has been clearly
established that in two-dimensions and zero temperature the ground state
is ``antiferromagnetic'' on a square lattice.  This means that the
``staggered'' spin-spin correlations between spins at distance r decay
to a nonzero constant as r grows.  Here a modulating sign +1 or -1 for
even and odd sites, respectively, is used in the definition of the
staggered spin correlation.  The decay to a finite constant at large
distances indicates that the ground state has long-range order in the
spin channel. In one-dimension (1D) quantum fluctuations are
sufficiently strong to prevent such a long-range order, but the
staggered spin-spin correlations decay slowly to zero as a power-law and
they remain dominant. Both in 1D and 2D there is no spin-gap, namely
there is no cost in energy to create a spin excitation with S=1.

The current interest in the study of ladder systems started when
Dagotto, Riera and Scalapino (1992) found that the S=1/2 Heisenberg
model defined now on a two-leg ladder instead of a chain or a plane, has
a $finite$ spin-gap in its spectrum of spin excitations (see also
Dagotto and Moreo, 1988). The reason leading to this result is simple in
the ``strong coupling'' or ``strong rung'' limit, namely the limit where
the Heisenberg coupling along the rungs ($\rm J_{\perp}$) is much larger
than along the legs ($\rm J$). In this case, the ground state of the
two-leg Heisenberg model corresponds to the direct product of
spin-singlets, one per rung, as schematically represented in Fig.1a.
The overall spin of the system is 0, since each rung pair of spins is
itself in a singlet. In order to produce a spin excitation, a rung
singlet must be promoted to a rung triplet, and this costs an energy
$\rm J_{\perp}$. These local excitations can propagate along the ladder
and its energy acquires a momentum dependence (see Barnes et al., 1993).
The spin-gap is the minimum at momentum $\pi$. Then, in the
strong-coupling limit the spin are mostly uncorrelated, being in
singlets most of the time, and the spin correlations decay with distance
exponentially along the chains. Such a state is usually referred to as a
``spin-liquid''. Gopalan, Rice and Sigrist (1994) and Noack, White and
Scalapino (1994) suggested that a good variational description of the
ground state can be obtained if the short-range resonance valence bond
(RVB) state (Kivelson, Rokhsar and Sethna, 1987) is used. Note that this
description is certainly accurate in the strong coupling limit, but in
the isotropic limit it must be supplemented by spin-singlets among spins
at distances larger than one lattice spacing to account for the observed
correlation length of about three lattice spacings, obtained from the
staggered spin-spin correlations (Noack, White and Scalapino, 1994).
Nevertheless, the short-range RVB-state is certainly a useful state to
use in qualitative considerations of the physics of ladders.  Other
studies of ladders based on the RVB picture have also been presented by
Sierra et al. (1998).

In the other limit $\rm J_{\perp} = 0$ the legs decouple and, as
remarked above, it is well-known that isolated chains do not have a
spin-gap and arbitrarily low-energy spin excitations can be created.
However, these chains are in a ``critical'' state since they do not
present long-range order but a power-law decay of correlations.  It is
anticipated that small perturbations may change qualitatively the
properties of its ground state. Based on these considerations, Barnes et
al. (1993) conjectured that a spin-gap opens on the two-leg ladders
immediately upon the introduction of a nonzero $\rm
J_{\perp}$. Numerical results supported this proposal and the spin-gap
was observed to be $\rm \sim J/2$ in the ``isotropic'' limit $\rm J =
J_{\perp}$, as also found numerically by Dagotto, Riera and Scalapino
(1992).  The actual spin-gap vs $\rm J_{\perp}/J$ is shown in Fig.2,
using the frequently applied alternative notation $\rm J_{\perp} = J'$.
Similar results were obtained by White, Noack and Scalapino (1994). In
addition, a spin-gap appears in the one-band Hubbard model at
half-filling if defined on a two-leg ladder (see Noack, White and
Scalapino, 1994; Azzouz, Chen and Moukouri, 1994). This result is
reasonable since the Heisenberg model is recovered from the
strong-coupling $\rm U/t$ limit of the one-band Hubbard model.  Since
the spin-gap is nonzero at all couplings with the only exception of $\rm
J_{\perp} = 0$, the ``physics'' of the two-leg ladders is said to be
dominated by the strong-coupling limit with nearly decoupled rungs.
Calculations in that regime are frequently carried out with the
expectation that the results will not change qualitatively in the
isotropic limit, which is more difficult to study directly due to the
absence of a small perturbative parameter.

If the number of legs increases in the $n$-leg ladder structure, the
physics of the two-leg ladders with its spin-gap should remain
qualitatively the same as long as $n$ is $even$. In this case, the
strong coupling limit $\rm J_{\perp} \gg J$ leads once again to the
formation of spin-singlets along the rungs since the ground state of $n$
(even) coupled S=1/2 spins has zero total spin. As a consequence, the
excitation of S=1 states costs a finite amount of energy. The gap must
decrease with $n$ such that the limit of a two-dimensional gapless plane
is recovered as $n$ grows, but its magnitude will remain nonzero for any
finite $n$ (even). Numerical calculations for the four-leg ladder by
Dagotto and Moreo (1988); Poilblanc, Tsunetsugu and Rice (1994); and
White, Noack and Scalapino (1994) support this picture. However,
different is the situation is $n$ is $odd$. If the strong coupling limit
is studied, in each rung an odd number of spins must be considered and
it is known that its ground state corresponds to a state with total spin
S=1/2. Then, at large $\rm J_{\perp}$ the system can be mapped into a
S=1/2 one-dimensional Heisenberg model with a coupling among spins
induced by a finite leg-coupling $\rm J$. This system is gapless. As a
consequence, these theoretical considerations, which were originally
proposed by Rice, Gopalan and Sigrist (1993) and Gopalan, Rice, and
Sigrist (1994), predict that odd-leg ladders should be gapless, while
even-leg ladders have a finite spin-gap as discussed above (see also
Rojo (1996)). Numerical calculations by White, Noack and Scalapino
(1994) are in agreement with this prediction. Using a Monte Carlo loop
algorithm and calculating the magnetic susceptibility as a function of
temperature, Frischmuth, Ammon and Troyer (1996) also found evidence
supporting the gapful vs gapless characteristics of the ground state of
even- and odd-leg ladders, respectively. Their results are reproduced in
Fig.~3.

\vspace{.6cm}
\centerline{\bf 2.2 Hole-doped Spin Ladder Models}
\vspace{0.6cm}

The theoretical study of ladder models in the presence of hole doping
leads to interesting predictions. Dagotto, Riera and Scalapino (1992)
studied the t-J model on a two-leg ladder using computational
techniques.  In this model the degrees of freedom at each site are
either a S=1/2 spin that mimics the $\rm Cu^{2+}$ ion with 9 electrons
in its d-shell or a ``hole'' that represents a $\rm Cu^{3+}$ state,
usually assumed to be made out of a spin-singlet combination of an
oxygen hole and the S=1/2 copper-ion forming a so-called Zhang-Rice
singlet. Analyzing the t-J model correlations corresponding to the
operator that create or destroy hole pairs at nearest-neighbor sites and
working at a small concentration of doped holes, Dagotto, Riera and
Scalapino (1992) observed that these correlations were very robust,
clearly indicative of a ground state dominated by strong superconducting
tendencies. The rationalization of this surprising result is sketched in
Fig.1b-c. The main idea is that the state upon which the holes are added
is dominated by the formation of rung singlets, at least in
strong-coupling. If a hole is added, equivalent to removing a S=1/2, the
other spin of the original singlet becomes free and it no longer reduces
its energy by singlet formation.  If two holes at large distance are
added to the system, each one will be producing a substantial energy
damage to the spin background as sketched in Fig.1b since both break a
singlet. However, if the two holes are nearby they can share a common
rung, reducing the number of damaged spin singlets from two to one. This
idea leads in a natural way to the concept of hole binding on two-leg
ladders (Fig.1c). Once the formation of hole pairs is accepted as a
logic consequence of the spin-singlet dominated characteristics of the
undoped ground state, the presence of superconductivity becomes natural
as explained by Dagotto, Riera, and Scalapino (1992) (see also Dagotto
and Rice, 1996).

Sigrist, Rice and Zhang (1994) using a mean-field approximation
predicted that the above described superconductivity in two-leg ladders
should exist in the ``d-wave'' channel, namely the pair correlations
should change sign if the creation (or destruction) hole-pair operator
is aligned along the legs or along the rungs. Although the two-leg
ladder is certainly not invariant under rotations as the planes are, the
language of referring to the superconductivity in ladders as ``d-wave''
is based on the similarities (sign change under rotations in 90$^o$)
with the same concept used on square-lattices. Actually since the
antiferromagnetic correlation length on ladders is about three lattice
spacings, which is sufficiently large for the individual holes to be
mainly surrounded by spins in an antiparallel arrangement, some of the
arguments that lead to d-wave pairing in the 2D cuprates (see for
instance Nazarenko et al., 1996) are also operative on ladders.  This
d-wave character of pairing on ladders has been confirmed by a variety
of calculations performed by Riera (1994), Tsunetsugu, Troyer and Rice
(1995), Hayward et al. (1995), and others.  Some results are shown in
Fig.4.  Calculations in the one-band Hubbard model arrived to similar
conclusions (see Asai (1994), Yamaji and Shimoi (1994)).  Additional
information about hole binding can be found in Gayen and Bose (1995).
Superconducting correlations have also been studied in three-leg ladders
(Kimura, Kuroki, and Aoki, 1996a; Kimura, Kuroki and Aoki, 1996b).  Very
recently, Gazza et al. (1999) analyzed the hole binding and pairing
correlations on two-leg ladders taking into account the Coulomb
repulsion between carriers at distances of one lattice-spacing which is
expected to be non-negligible. This interaction is rarely considered in
the study of the t-J model.  Gazza et al. (1999) found that in spite of
this repulsion the binding and pair correlations are still robust on
ladders, as it occurs when only on-site Coulomb interactions are
included. This provides additional support to the idea that real doped
ladders should be superconducting.

It is likely that these superconducting tendencies will be robust as
long as the spin background does not alter drastically its spin-gap
properties. However, it is also expected that the spin-gap will reduce
its size as the hole doping increases (see Dagotto, Riera, and
Scalapino, 1992; Noack, White and Scalapino, 1994; Poilblanc, Tsunetsugu
and Rice, 1994; Hayward and Poilblanc, 1996). In this context, the spin
excitations created upon hole doping were analyzed by Tsunetsugu, Troyer
and Rice (1994). Besides the excitation that already appears in the
undoped limit, namely the promotion of a rung spin-singlet to a triplet,
other possible excitations may arise. For instance the hole pairs
described in the previous paragraph may change their spin-singlet
characteristics to a triplet. In other words, a bound state of a
two-hole bound state and a rung spin-triplet may develop. The existence
of this excitation has been recently confirmed by Dagotto et al. (1998)
using a new technique that uses just a fraction of the total Hilbert
space of a given cluster (Riera and Dagotto, 1993, and references
therein), after the Hamiltonian is rewritten in the basis that
diagonalizes the individual rungs.  The undoped limit has also been
studied in the plaquette-basis, instead of the rung-basis, as reported
by Piekarewicz and Shepard (1997).  A typical result in the rung-basis
is shown in Fig.5, working on a $2 \times 16 $ cluster. Of the two
branches, one (I) is a remnant of the result observed without holes and
the other one (II) is generated by hole doping and it corresponds to the
hole-pair rung-triplet bound state. Other studies using the rung-basis
have involved the calculation of the one-particle spectral function, and
the results will be briefly summarized when photoemission experiments
are discussed later in the review.

The list of interesting theoretical studies of ladder systems is
certainly much longer than shown thus far in this section. Other aspects
of ladder systems that have been analyzed in the literature include the
careful study of antiferromagnetic correlation lengths vs temperature in
the undoped limit (see Greven, Birgeneau and Wiese, 1996; Syljuasen,
Chakravarty and Greven, 1997), weak coupling renormalization-group
treatments (Kuroki and Aoki, 1994; Balents and Fisher, 1996a; Lin,
Balents and Fisher, 1997; Lin, Balents and Fisher, 1998; Emery, Kivelson
and Zachar, 1998), coupling of the chains with emphasis on the Luttinger
liquid properties (Schulz, 1996), the influence of four-spin
interactions in the undoped ground state (Nersesyan and Tsvelik, 1997),
staggered spin ladders (Martin-Delgado, Shankar and Sierra, 1996),
influence of disorder on ladders (Orignac and Giamarchi, 1998), two-leg
Hubbard ladders and its relation with carbon nanotubes (Balents and
Fisher, 1996b; Konik et al., 1998) and comparison with weak-coupling
calculations (Park, Liang and Lee, 1998), influence of interladder
coupling on the possible dimensional crossover of ${\rm Sr_{14-x} Ca_{x}
Cu_{24} O_{41} }$ (Kishine and Yonemitsu, 1998), the relation between
the resistivity of this material and the two-dimensional $\rm Y Ba_2
Cu_4 O_8$ cuprate (Moshchalkov, Trappeniers and Vanacken, 1998), as well
as several others.  In addition, ladder physics is also considered in
the context of stripe phases for the two-dimensional cuprates since the
region in between the presumed to exist one-dimensional-like metallic
stripes are $n$-leg ladders. The spin-gap proximity effect discussed by
Emery, Kivelson and Zachar (1997) proposes that a spin-gap is created in
the 1D metallic striped environment due to the influence of a nearby
system with spin-gapped properties, such as an even-leg ladder.  The
list of interesting ladder-related theoretical results goes on and on.
However, since the focus of this review is on the experimental results
found on ladder materials rather on theoretical developments, the author
will postpone a comprehensive review of the latter for the near future
and concentrate the efforts on the experimental aspects of ladder
physics in the rest of the paper.

\section{Experimental Results}

After an enormous experimental effort, reviewed here, several materials
with ladders in their structures have become available in recent years,
and a variety of exciting experiments have already been carried out in
this context. In the present section the main experimental results will
be discussed, organizing the presentation according to the family of
ladder compound studied, and the technique used in its analysis. The
main emphasis will be given to the Cu-oxide based ladders since they
admit hole-doping by chemical substitution. Among them
Sr$_{14-x}$Ca$_{x}$Cu$_{24}$O$_{41}$, the so-called ``14-24-41''
compound, will be discussed in particular detail since superconductivity
has been found in this material, as described below.
Note, however, that the compound that was mentioned in the early work of
Dagotto, Riera and Scalapino (1992) as a possible realization of a
ladder model was of a different variety, namely $\rm (VO)_2 P_2 O_7$
(sometimes known as VOPO). This material, originally discussed by
Johnston et al. (1987), has a two-leg V-oxide ladder in its
structure. Neutron scattering experiments on VOPO presented by Eccleston
et al. (1994) reported a spin-gap in the spectra which was interpreted
in terms of a two-leg ladder dominant structure.  However, more recent
inelastic neutron scattering experiments by Garrett et al. (1997a) have
shown that the properties of this compound are better described by an
alternating Heisenberg antiferromagnetic chain, running perpendicular to
the ladders (see also Garrett et al., 1997b).  Based on this reanalysis,
it is reasonable at present to assume that VOPO is no longer part of the
family of ladder compounds, and thus the description below will be
concentrated on other materials, notably those with chemical
compositions containing copper and oxygen, similar to the high-$\rm T_c$
cuprates, since they are susceptible to chemical substitutions and the
concomitant addition of hole carriers.

\vspace{0.6cm}
\centerline{\bf 3.1 The Cu-oxide Ladder Compound SrCu$_2$O$_3$}
\vspace{0.6cm} 

A seminal contribution to the development of the field of ladders was
provided by Azuma et al. (1994) (see also Hiroi et al. (1991)), when
these authors reported the preparation of a copper-oxide S=1/2 two-leg
ladder. This result was not only important for providing a concrete
realization of the ladder compounds discussed by the theorists, but also
due to the possibility of changing the hole concentration through
chemical substitution in such a Cu-oxide, as it occurs in the high-$\rm
T_c$ cuprates.  The chemical composition of the compound reported by
Azuma et al. (1994) was $\rm Sr Cu_2 O_3$, and a schematic
representation is provided in Fig.~6. Note that within the ladders the
180$^o$ Cu-O-Cu bond should provide a strong Heisenberg coupling among
the Cu-ions, while in between the ladders the bond is of 90$^o$ and,
thus, expected to be weak and ferromagnetic.

In Fig.~7 the experimentally determined temperature dependence of the
magnetic susceptibility of $\rm Sr Cu_2 O_3$ is shown (from Azuma et
al., 1994). The open circles are the experimental raw numbers, while as
closed circles it is shown the data after subtraction of the Curie
component due to impurities.  The solid line through them is a
theoretical fit that gives a spin-gap of 420K. This is in rough
agreement with the theoretical expectation of $\rm \Delta \sim J/2$
reported by Barnes et al. (1993), if it is assumed that the Heisenberg
coupling is similar in magnitude to its two-dimensional copper-oxide
counterpart $\rm J \sim 1200K$.  This is in principle a natural
assumption due to the similarities between the Cu-O-Cu bonds on ladders
and planes. Note that Azuma et al. (1994) also observed (Fig.8) that the
magnetic susceptibility of a three-leg ladder material ($\rm Sr_2 Cu_3
O_5$) does not show indications of a spin-gap, as expected from a ladder
with an odd number of legs (see Rice, Gopalan, and Sigrist, 1993;
Gopalan, Rice and Sigrist, 1994; Sigrist, Rice, and Zhang, 1994; Dagotto
and Rice, 1996). Muon spin relaxation measurements of the two- and
three-leg compounds by Kojima et al. (1995) have also confirmed the
gapful vs gapless character of the spin state of even- and odd-leg
ladder systems. Recent experiments by Thurber et al. (1999) suggest that
the three-leg ladder compound has a dimensional crossover from quasi-1D
to anisotropic-2D at 300K.

$\rm ^{63}Cu$ Nuclear Magnetic Resonance (NMR) studies by Ishida et
al. (1994) and Ishida et al. (1996) also contributed to the analysis of
the two- and three-leg ladder compounds described here.  The spin gap of
$\rm Sr Cu_2 O_3$ was in this case found to be 680K, i.e. larger than
the result deduced from the magnetic susceptibility by Azuma et
al. (1994), although of the same order of magnitude. The instantaneous
spin-spin correlation was estimated to be 3-4 lattice spacings, in good
agreement with theoretical expectations (White, Noack, and Scalapino,
1994, Dagotto and Rice, 1996, and references therein).

Regarding the difference between the spin-gap reported by magnetic
susceptibility and NMR techniques, as well as the extra difficulty in
understanding the origin of the actual value of the exchange Heisenberg
constant which seems smaller than in other 180$^o$ Cu-O-Cu bond
materials, Johnston (1996) made the important observation that the
susceptibility data of $\rm Sr Cu_2 O_3$ can be fit accurately with a
two-leg ladder prediction using a rung coupling, denoted by $\rm J'$ in
Fig.9, smaller by a factor two than the coupling along the legs J (see
Barnes and Riera, 1994).  Using the ratio $\rm J'/J \sim 0.5$, the
exchange constant needed to fit the data is now $\rm J \sim 2000K$ in
agreement with results obtained for other cuprates by Motoyama, Eisaki
and Uchida (1996) (for details see Johnston, 1996). Then, although the
microscopic origin of the anisotropy between legs and rungs is still
under discussion, evidence is accumulating that this compound
corresponds to a ladder in the coupling regime $\rm J'/J <1$ (note,
however, that a ratio $\rm J'/J$ very close to 1 has been obtained by
Matsuda et al. (1999) by including four-spin exchange interactions in
the ladder model Hamiltonian to fit neutron scattering data for $\rm
La_6 Ca_8 Cu_{24} O_{41}$). It is unclear to this author if for values
of $\rm J'/J$ smaller than one superconductivity would be expected upon
hole doping for realistic values of $\rm J/t$ in the t-J model discussed
in the previous section. Calculations should be carried out in the
interesting regime $\rm J'/J \sim 0.5$ to address this issue. Adding to
the analysis of this material, Ishida et al. (1996) reported the effect
of doping carriers on the two-leg compound by substitution of Sr by
La. Their results suggest that the added electrons unfortunately remain
localized, and thus electron doped $\rm Sr Cu_2 O_3$ may not provide a
good realization of the physics arising from the (hole) doping of the
t-J model. Other compounds are likely needed to realize the theoretical
scenario discussed in Section 2.2.

\vspace{0.6cm}
\centerline{\it 3.1.1 Doping of SrCu$_2$O$_3$ with Zinc}
\vspace{0.6cm} 

The doping of $\rm Sr Cu_2 O_3$ through the replacement of $\rm Cu^{2+}$
by $\rm Zn^{2+}$ has produced interesting results. Particularly
important has been the observation by Azuma et al. (1997) of a rapid
suppression of the spin-gap in this two-leg ladder compound as the
concentration of Zn (here denoted by x) was increased. The results for
the magnetic susceptibility are shown in Fig.10. As x increases, the
susceptibility also increases at low-temperature revealing the presence
of low-energy spin states.  At those temperatures an unexpected
transition to an antiferromagnetic state is also observed, with a
maximum N\'eel temperature of 8K. Then, it is apparent that a very small
amount of impurities affects severely the low-temperature properties of
${\rm Sr Cu_2 O_3 }$ (see also Fujiwara et al., 1998).  Recently, Ohsugi
et al. (1998) reported a comprehensive study of $\rm Sr Cu_2 O_3$ doped
with Zn, Ni and La. The results were analyzed in terms of an
impurity-induced staggered polarization around the vacancies. The size
of this perturbation centered at, e.g., the Zn-ion was found to be very
large, much longer than the 3-5 lattice spacings characteristic of the
antiferromagnetic correlation length without the impurities. Actually
the new correlation length is as large as 50 lattice spacings at
Zn-doping x=0.001. Although rather surprising note that these results
are qualitatively similar to those reported by Laukamp et al. (1998)
using computational techniques in their study of the influence of
vacancies on the spin arrangement in its vicinity, using two-leg ladders
and neglecting interladder couplings.  For Ni-doping similar theoretical
results were obtained by Hansen et al. (1998).  It is interesting to
observe that a quite similar behavior has been found in spin-Peierls
chains, which also have a spin-gap although of an origin quite different
from the one found in ladders (for references on experiments and
theoretical developments in this context see Laukamp et al., 1998).  A
possible rationalization of this curious phenomenon is the following: in
the global spin-singlet state of the undoped ladder, it is known that
the formation of local spin-singlets along the rungs is favored in the
sense that such a configuration has a substantial weight in that ground
state. If one of the spins is replaced by a zero-spin vacancy, such as
$\rm Zn^{2+}$, the other spin belonging to the same rung becomes
``free'' and, thus, it contributes to the magnetic susceptibility at
low-temperatures as a S=1/2 impurity would do. These impurities are
certainly weakly interacting among themselves since their mean distance
is large, but at very low-temperature such a residual interaction may
nevertheless favor a N\'eel order state, as discussed by Martins et
al. (1996), Martins et al. (1997) and Laukamp et al. (1998). The reader
should also consult the related literature contained in the work of
Motome et al. (1996), Ng (1996), Nagaosa et al. (1996), Miyazaki et
al. (1997), and references therein.  Some of these authors also
discussed the related $enhancement$ of spin correlations in the vicinity
of the vacancies, as observed in computational studies of spin systems
for both ladders and dimerized chains.  This enhancement favors the
tendency towards antiferromagnetic order as the concentration of
vacancies grows. As a consequence, from an apparently disordering effect
such as the random replacement of Cu-ions by Zn-ions, an order state is
stabilized at low-temperatures.
 
Laukamp et al. (1998) have also analyzed the ratio $\rm J_{\perp}/J$
that fits the data the best, and, in agreement with Johnston (1996), it
was observed that a ratio 0.5 leads to an antiferromagnetic disturbance
near the vacancies that covers the entire lattice even at very low
Zn-concentrations.  As example, the (staggered) local susceptibility
calculated using numerical techniques is shown in Fig.11 for three
values of $\rm J_{\perp}/J$, namely 0.5, 1.0 and 5.28, the latter
corresponding to the organic compound ${\rm Cu_2(C_5 H_{12} N_2 )_2
Cl_4}$, which will be discussed in another section of this review.
Clearly at 0.5 is that all spins become involved in the staggered spin
order at this vacancy density, a result compatible with experiments. The
results in Fig.11 are actually very similar to the data in Fig.13 of
Ohsugi et al.  (1998) that reported the spatial variation of the
antiferromagnetic moments in 1\% and 2\% Zn-doped $\rm Sr Cu_2
O_3$. Then, theoretical studies are able to reproduce the qualitative
features of the experiments in ladders with vacancies.

\vspace{0.6cm}
\centerline{\bf 3.2 The metallic ladder compound La$_{1-x}$Sr$_x$CuO$_{2.5}$}
\vspace{0.6cm}

Under a high pressure of 6~GPa, Hiroi and Takano (1995) made the
important discovery that the compound $\rm La_2 Cu_2 O_5$ undergoes a
transformation into a simple perovskite structure. The ionic arrangement
of this compound is shown in Fig.12.  Two-leg copper-oxide ladders
appear in the structure, involving 180$^o$ angle Cu-O-Cu bonds. The
length and angle of the corresponding Cu-Cu bonds between ladders
suggest that the interladder coupling is weaker than the intraladder
one. Then, as a first approximation, it is reasonable to assume that
this compound is formed by an array of weakly interacting two-leg
ladders.  More details about this material and the experiments can be
found in the comprehensive review by Hiroi (1996).  The temperature
dependence of the magnetic susceptibility of $\rm La Cu O_{2.5}$ is
shown in Fig.13, reproduced from Hiroi (1996). After removing the
contribution of impurities, the resulting susceptibility indeed behaves
as expected for a two-leg Heisenberg ladder. In particular, it presents
a spin-gap behavior at low-temperatures with $\Delta = 492K$, a number
similar to the gap obtained for $\rm Sr Cu_2 O_3$ ($\sim 400K$), as
described in the previous sections.

An important observation is that the ladder compound $\rm La Cu O_{2.5}$
can be hole doped replacing La by Sr (Hiroi, 1996).  This doping is
achieved also working at high pressure. There are no significant changes
in the structure after doping.  The amount of Sr that can be added to
$\rm La Cu O_{2.5}$ is fairly large, up to x=0.20. Following this doping
procedure a dramatic insulator-to-metal transition has been observed in
the resistivity, as shown in Fig.14, taken from Hiroi (1996).  Based on
the sign of the slope of the resistivity vs temperature, the
metal-insulator transition was found to be located roughly between
x=0.18 and 0.20. Note the sharp decrease in the resistivity with doping
which corresponds to more than seven orders of magnitude. Comparing the
results for this compound against those for the two-dimensional cuprate
$\rm La_{2-x} Sr_x Cu O_4$, it was observed that at the hole density
$\rm x \sim 0.20$, the resistivities of both compounds are actually very
similar, as the reader can observe in Fig.15 (Hiroi, 1996).  The doping
dependence of the magnetic susceptibility is contained in Fig.16.  The
spin-gap characteristics of the ground state seem to disappear with
increasing doping.  Photoemission and x-ray-absorption studies performed
on ${\rm La_{1-x} Sr_x Cu O_{2.5}}$ by Mizokawa et al. (1997) found
changes in the overall electronic structure induced by hole doping
similar to those in $\rm La_{2-x} Sr_x Cu O_4$. Spectral weight around
the Fermi level ($\rm E_F$) was found to increase with x, and a weak but
finite Fermi edge is established at $x=0.20$. This result is compatible
with the metallic character of the compound deduced from the resistivity
at this hole concentration.  Then, it is clear that ${\rm La_{1-x} Sr_x
Cu O_{2.5}}$ is a metal at $\rm x$$\sim$$0.20$, improving on the
characteristics of $\rm Sr Cu_2 O_3$ (Sec. 3.1) which is a two-leg
ladder compound that cannot be hole doped.

Although the resistivity results presented in Fig.14 correspond to an
encouraging metal-insulator transition upon doping, unfortunately those
results also show that this ladder compound does not become a
superconductor in spite of the theoretical predictions in this
respect. There are two possible reasons to understand the absence of
superconductivity here. First, the interladder coupling, while certainly
smaller than the intraladder one, may not be negligible small since it
does not involve 90$^o$ bonds.  It is not clear to what extend this
finite extra interladder coupling would change the theoretical
predictions for isolated ladders. Second, the replacement of La by Sr
produces an intrinsic randomness in the system that may localize the
holes.

Regarding the first possibility, namely the relevance of the interladder
couplings in this compound, Normand and Rice (1996) proposed that the
magnetic state of $\rm La Cu O_{2.5}$ may be located in parameter space
near the crossover from spin-liquid to the antiferromagnetic state
expected on a three-dimensional spin system without frustration (see
also Normand and Rice (1997)).  This proposal is motivated by the NMR
results of Matsumoto et al. (1996) and $\mu$SR results of Kadono et
al. (1996) that actually reported the presence of antiferromagnetic
order in this compound at a temperature $\rm T_N \sim 110K$. Then, it is
apparent that the spin-singlet state is in close competition with a
N\'eel state in this compound.  Troyer, Zhitomirsky and Ueda (1997),
using a combination of analytical and quantum Monte Carlo techniques,
expanded on Normand and Rice's idea and showed that the apparently
conflicting experimental results, namely evidence of spin-liquid
formation at intermediate temperatures and a N\'eel state at
low-temperature, can be reconciled if the system is near a quantum
critical point. However, more recently Normand, Agterberg and Rice
(1998) argued that even including realistic interladder couplings, the
doped $\rm La Cu O_{2.5}$ compound should present superconductivity, at
least within the spin-fluctuation approximation. Then, these authors
concluded that the random potential due to the replacement of $\rm
Sr^{2+}$ by $\rm La^{3+}$ must be responsible for suppressing the
superconducting phase, and suggested that more work should be devoted to
the preparation of single crystals of this compound at low-doping values
to search for indications of superconductivity, which is expected to be
the strongest in such a regime of density according to their theoretical
calculations. Overall, it is clear that the experimental analysis of the
properties of this ladder compound should continue in order to clarify
what causes the suppression of superconductivity, and to investigate
what sort of metallic state is obtained near the spin-gapped regime that
resembles the behavior of the 2D cuprates.

\vspace{0.6cm}
\centerline{\bf 3.3 The Superconducting Ladder Compound} 
\centerline{\bf Sr$_{14-x}$Ca$_{x}$Cu$_{24}$O$_{41}$}
\vspace{0.6cm}

After the quite interesting first attempts described in Sections 3.1 and
3.2 to search for superconductivity in two-leg ladders, success was
finally reached using the compound $\rm (La,Sr,Ca)_{14} Cu_{24} O_{41}$,
the 14-24-41 or ``phone number'' compound. The synthesis of this
material was initially reported by Mc Carron et al. (1988) and Siegrist
et al. (1988). This compound contains 1D-CuO$_2$ chains, (Sr,Ca) layers,
and two-leg $\rm Cu_2 O_3$ ladders as shown in Fig.17. A very important
property of this material is that it can be synthesized at ambient
pressure, contrary to some of the compounds described in previous
Sections.  In addition, a spin-gap was confirmed to exist in the $\rm
Sr_{14} Cu_{24} O_{41}$ compound as shown, for example, using inelastic
neutron scattering by Eccleston, Azuma, and Takano (1996). The nominal
Cu valence for this material is +2.25, which arises from $\rm [2 \times
41(Oxygens) - 2 \times 14(Sr)]/24(Cu)$, instead of just +2.0.  This
means that holes are already doped into the structure (``self-doped''
system), similarly as it occurs in more familiar high-$\rm T_c$
compounds such as $\rm YBa_2 Cu_3 O_7$.  Replacement of Sr by Na, which
is in a 1+ state, is possible and NMR studies have been recently
reported for this compound by Carretta, Ghigna, and Lascialfari (1998).
However, in practice the substitution of Sr by Ca, both of them in a 2+
valence state, turned out to be more efficient in providing carriers to
the ladders since such a doping may alter the distribution between the
chains and ladders of the already existing holes.  Carter et al. (1996)
in their important early investigations in this context indeed managed
to replace Sr by Ca increasing by this procedure the number of hole
carriers on the ladders-chains structure. These authors reported a
decrease of the resistivity by this procedure, and interpreted their
results as corresponding to the addition of extra holes to the chains
instead of the ladders.  However, the prevailing current point of view
is that Ca-doping produces a rearrangement of holes, moving them into
the ladders, as explained in detail in the subsequent sections.  The
discussion that follows on the notable 14-24-41 compound will be
organized starting with the remarkable discovery of superconductivity,
followed by a review of the results obtained in this context using a
variety of experimental techniques.

\vspace{0.6cm} 
\centerline{\it 3.3.1 Superconductivity in Doped 14-24-41 with x=13.6}
\vspace{0.6cm} 

A crucial breakthrough in the field of ladders occurred when Uehara et
al.  (1996) discovered superconductivity in the x=13.6 Ca-doped compound
$\rm Sr_{0.4} Ca_{13.6} Cu_{24} O_{41.84}$.  To stabilize
superconductivity Uehara et al. (1996) observed that a high pressure of
3 GPa is needed.  This discovery substantially contributed to the
current enormous effort devoted to the understanding of ladder compounds
(Levy, 1996).
The resistivity $\rho$ vs temperature at several pressures is shown in
Fig.18.  The drop in $\rho$ at the critical temperatures $\rm T_c$ is
clear.  There is an ``optimal'' pressure at which the $\rm T_c$ is the
highest.  More recent studies of $\rm Sr_{0.4} Ca_{13.6} Cu_{24} O_{41 +
\delta}$ by Isobe et al. (1998) reported the resistivity vs temperature
shown in Fig.19. The results are very similar to those previously
reported by Uehara et al. (1996). The hump centered at $\sim 100K$
disappears at 5~GPa and the sample goes into a metallic regime.  Note
that the resistivity at x=13.6 and high pressure changes from an
approximate $\rm T^2$-dependence near the optimal pressure to a linear
dependence at higher pressures when superconductivity is suppressed (see
Isobe et al., 1998). This is different from the behavior observed in the
2D high-$\rm T_c$ compounds and also different from observation at other
Ca concentrations.  The critical temperature vs pressure from Isobe et
al. (1998) is reproduced in Fig.20.  The bell-like shape of the curve
clearly resembles the well-known results for the 2D cuprates, if it is
assumed that pressure is replaced by hole concentration. The optimal
pressure is around 5~GPa, with an optimal $\rm T_c$ close to 14K.  While
this critical temperature is obviously lower than those reached in
two-dimensional high-$\rm T_c$ cuprates, it is nevertheless higher than
those of typical metallic superconductors in spite of its low-carrier
density.  In addition, it is important to remark that this compound is
the only known copper-oxide superconductor $without$ two-dimensional
planes. Then, independently of whether this material provides or not a
realization of the theoretical ideas based on two-leg ladders reviewed
in Sec.2, it nevertheless provides a very interesting case for the
analysis of superconductivity in copper-oxides in general.

After the report of superconductivity in doped 14-24-41, it became
very important to verify that the structure at ambient pressure with
clearly defined ladders and chains remains stable under the high
pressure needed to induce the superconducting regime.  In other words,
pressure-induced structural changes could have occurred in the
14-24-41 material. This issue has been addressed recently by Isobe et
al. (1998) using x-ray diffraction measurements on a x=13.6 Ca-doped
sample. Among the main results reported by these authors is the
confirmation that indeed no serious structural changes take place in
this compound under high pressure, and as a consequence it is safe to
consider that the superconducting phase is related with the original
ladder-chain structure observed at ambient pressure. The main effect of
pressure apparently is to reduce the distance between the ladders and
chains ($b$ direction), as the evolution of the lattice constants with
pressure shown in Fig.21 suggests. This result is representative of hard
intraplane and soft interplane bindings. A similar conclusion is reached
by analyzing the effect of Ca-doping upon the original Sr-based
Ca-undoped phase, namely the interatomic distance ladder-chain was found
to be reduced by Ca-substitution (see Ohta et al., 1997) leading to a
redistribution of holes originally present only on the chains, as also
discussed by Kato et al. (1996), Motoyama et al. (1997), Osafune et
al. (1997), and Mizuno, Tohyama, and Maekawa (1997). According to Isobe
et al. (1998), the most important role of high pressure for realizing
superconductivity is the hole redistribution between chains and
ladders. In addition, the resistivity along the $a$-axis decreases
substantially with pressure enhancing the two-dimensional
characteristics of the compound, as discussed extensively below.

\vspace{0.6cm} 
\centerline{\it 3.3.2 Superconductivity in Doped 14-24-41 with x=11.5}
\vspace{0.6cm}

After the discovery of superconductivity at Ca-concentration x=13.6, a
similar phenomenon was also observed at x=11.5 by Nagata et al. (1997).
At 4.5 GPa the superconducting critical temperature at this
Ca-composition was found to be 6.5 K (see inset of Fig.22 for more
details).  Very recent specific heat and neutron scattering studies by
Nagata et al. (1998b,1999) have found that an antiferromagnetic phase
with a N\'eel temperature $\rm T_N \sim 2K$ exists at this
Ca-composition, both at low pressures and at pressures comparable to
those that induce superconductivity.  Then, in the temperature-pressure
diagram antiferromagnetism and superconductivity are neighboring
phases. The discussion of the complicated spin arrangement leading to
the magnetic ordering can be found in Nagata et al. (1998b,1999).
However, since $\rm T_c$ is appreciably larger than $\rm T_N$, it is
more reasonable to continue assuming that the superconducting phase is
more likely induced by the spin-gapped normal state that exists in these
ladder materials.

In Fig.22a, the resistivity along the $c$-direction (legs) is shown at
several pressures as a function of temperature.  Indications of
superconductivity below 10K have been found in the range from 3.5 GPa to
8GPa.  It is interesting to note the $linear$ T-dependence of $\rho_c$
above 80K in the ambient pressure limit, similarly as it occurs in the
2D high-$\rm T_c$ cuprates at optimal density.  This linear behavior can
also be observed in a wide range of pressures, but as the pressure
increases a crossover to a $\rm T^2$-dependence was found (see, for
instance, the inset of Fig.4 of Nagata et al., 1998a). Actually at 8~GPa
a Fermi-liquid $\rm T^2$-dependent resistivity was observed between 50 K
and 300K.
The behavior reported in this compound is, thus, very similar to what
occurs in the two-dimensional cuprates between the underdoped or optimal
and overdoped regions.  This similarity suggests that an increase of the
pressure may correspond to an increase in the number of carriers on the
ladder planes. The Fermi liquid behavior at large pressure in the region
where superconductivity disappears suggests that a 2D Fermi liquid state
is not favorable for superconductivity, as it occurs in the overdoped
regime of the two-dimensional high-$\rm T_c$ cuprates.

Nagata et al. (1997) also reported the resistivity $\rho_a$ in the
direction perpendicular to the legs, along the ladder rungs (Fig.22b). A
semiconducting-like behavior is observed below 1.5~GPa, suggesting that
the charge dynamics is essentially one-dimensional and carriers are
confined within each ladder.  However, above this pressure the system
becomes metallic. It is likely that the two-dimensional character of the
electronic state increases as the interladder coupling increases due to
the application of pressure. Even more illustrative is the ratio
$\rho_a/\rho_c$ vs temperature at various pressures shown in Fig.23,
reproduced from Nagata et al. (1997).  At low pressure this ratio can be
as large as 100, while at the pressure where the superconducting volume
fraction is at its maximum, $\rho_a/\rho_c$ reaches its minimum value
$\sim 10$. This result suggests that at low pressure the carriers are
confined on ladders, very similarly as holes are confined on the 2D
planes of the high-$\rm T_c$ cuprates, while as the pressures increases
they become deconfined. However, it is somewhat surprising that at the
highest pressure of 8~GPa, where the two-dimensional character should be
maximized following the previous reasoning, $\rho_a/\rho_c$ is now $\sim
20$ i.e. twice the minimum reached at 4.5~GPa.

It is certainly important to study the qualitative effect of pressure on
the physics of the ladder compound 14-24-41. In particular, it should
be clarified what is the role of the interladder coupling in the
appearance of superconductivity.  Arai and Tsunetsugu (1997) have
recently estimated the interladder hopping to be 5\% to 20\% of the
intraladder hopping. However, it is unclear whether this number should
be considered ``small'' or ``large''.  It could occur that under high
pressure the quasi-1D charge dynamics is preserved and superconductivity
develops mainly within the ladders, with a minor role played by the
interladder hopping amplitude. Another possibility is that the
interladder coupling, presumably enhanced as the pressure grows, induces
a dimensional crossover from 1D to 2D. In this scenario, it may simply
occur that the ladders become an anisotropic two-dimensional
copper-oxide system after the application of pressure.  These issues
have been recently addressed in more detail by Nagata et al. (1998a) and
the interpretation they provide of their results seem to favor the
second scenario, i.e. they consider that pressure makes the 14-24-41
compound quasi two-dimensional. They base their conclusion on the
$\rho_c$ and $\rho_a$ resistivities shown in Fig.23.  Along the
$c$-direction and low-temperature the system changes from an insulator
to a superconductor, similarly as observed in 2D cuprates upon Zn
substitution (see Fukuzumi et al., 1996). Along the $a$-direction,
$\rho_a$ becomes metallic at 4.5 GPa.

Regarding the results of Nagata et al. (1998a) this author would like to
bring to the attention of the readers similar results obtained
experimentally for the two-dimensional high-$\rm T_c$ compounds, in
particular $\rm Y Ba_2 Cu_3 O_{7-y}$. Takenaka et al. (1994) compared
the resistivity along one of the directions of the $\rm CuO$ planes
($a$-axis) against the resistivity along the direction perpendicular to
the planes, which in the high-$\rm T_c$ cuprates is the $c$-axis. The
results are shown in Fig.24. It is important to note that near the
$optimal$ critical temperature, where $\rm T_c$ is the highest, the
resistivity in $both$ directions $a$ and $c$ is metallic. Then, in this
regime the holes seem to have been deconfined from the planes, and they
can move coherently between planes. The anisotropy $\rm \rho_c/\rho_a$
of YBCO is shown in Fig.25.  Close to the optimal oxygen concentration
and at 100K this ratio is close to 50. Based on these numbers, it is
difficult to avoid making a connection with the results obtained for the
doped ladders. In that case at the optimal pressure around 4.0-4.5 GPa
the resistivity in both directions (along the legs and rungs of the
ladders) is metallic, and the ratio $\rm \rho_c/\rho_a$ is between 16
and 18. The difference between 50 for optimal YBCO and 16-18 for optimal
14-24-41 does not seem particularly large.  In this author's opinion,
the combination of these results is suggestive of an interesting common
phenomenon taking place at optimal hole concentration both in ladders
and planes. The best $\rm T_c$ is obtained in a regime where the charge
becomes deconfined to a dimension higher by one with respect to the
dimension of the dominant structure in the system, namely planes in the
2D cuprates and ladders in 14-24-41.

If this analogy is correct, this author believes that the doped ladder
compound should present an ``underpressure'' or underdoped regime, with
metallic behavior along the $c$-axis and semiconducting at
low-temperature along the $a$-axis. Based on the recent results of
Nagata et al. (1998a), such a regime could occur at pressures below 4.0
GPa.  According to Nagata et al. (1998a) indications of
superconductivity already occur at 3 GPa, and in this case the ratio
$\rho_a/\rho_c$ is 40, only about half the maximum reached at ambient
pressure. Although the here proposed underdoped regime would exist, if
at all, in a very narrow window in pressure, the dependence of the
ladder hole concentration with such a pressure is not accurately known
and it may occur that changing from 3.0 to 4.0 GPa alters appreciably
the number of carriers on the ladders. Of course, another possibility is
an abrupt change from the insulator to an ``overdoped'' superconductor
as it may happen in the 2D electron-doped NdCeCuO high-Tc compound
(S. Uchida, private communication).  The author believes that the study
of this conjectured underdoped regime would further contribute to the
understanding of the analogies and differences between the ladder and
planar cuprates.

Adding to this discussion, note that even if high pressure indeed
increases the two-dimensional characteristics of the system at all
Ca-concentrations where superconductivity is observed, still the x-ray
experiments of Isobe et al.  (1998) have shown that the structure
remains the same, namely with a weak interladder-coupling and with
nearest-neighbor ladders shifted by half a lattice spacing. This
geometry certainly does $not$ correspond to a square lattice, and in
addition it is likely that the dominant physics will still arise from
the individual ladders since they have the largest couplings and
$\rho_a/\rho_c$ much larger from one. In other words, if the
superconductivity on ladders and 2D cuprates have a common origin, any
mechanism proposed for the explanation of such a phase on the
square-lattice high-$\rm T_c$ cuprates cannot depend on the fine details
of such a lattice structure since ladder compounds under pressure are
also superconducting.
Also note that a competing charge-ordered state, with hole pairs as
building blocks, may affect the occurrence of superconductivity at
low-pressure, and only at high-pressure the extra mobility in the
direction perpendicular to the ladders stabilizes the superconducting
state, which still could occur through a pairing mechanism taking place
mostly on ladders.  Then, the issue of the 1D vs 2D characteristics of
the superconducting phase still is under much discussion, and this
crucial issue should certainly be addressed by future experiments.

Very recently, exciting results by Balakirev et al. (1998) have added
extra evidence of an unconventional behavior in doped ladders. These
authors showed that charge transport in the non-superconducting state of
$\rm Sr_2 Ca_{12} Cu_{24} O_{41}$ shares three distinct regimes in
common with high temperature superconductors, including an unexplained
insulating behavior at low-temperatures in which the resistivity
increases as the logarithm of the temperature. A similar regime was
found previously in the 2D cuprates by Ando et al. (1995). The other two
regimes correspond to a variable-range-hopping behavior at very
low-temperature and a linear behavior above 150K, as already discussed
elsewhere in the text. Boebinger et al. (1996) studying the high-$\rm
T_c$ superconductors in pulsed high magnetic fields found an insulator
to metal crossover at low-temperature in $\rm La_{2-x} Sr_x Cu O_4$. The
suppression of superconductivity by the magnetic fields allowed them to
analyze this insulator-metal transition below $\rm T_c$.  The study of
the analog metal-insulator transition in magnetic fields strong enough
to suppress superconductivity for the case of two-leg ladders 14-24-41
at high pressure would be an important experiment to carry out in the
near future. Its main motivation would be to find out if such a
transition occurs at the optimal pressure, similarly as it happens in
the 2D cuprates at optimal hole-density.

\vspace{0.6cm}
\centerline{\it 3.3.3 Optical Conductivity $\sigma(\omega)$:}
\vspace{0.6cm}

The study of the possible transfer of holes from chains to ladders upon
Ca-doping has been illuminated by experiments reporting the optical
conductivity along the $c$-axis carried out by Osafune et al. (1997).
They used single crystals of doped 14-24-41 with Ca-concentration up
to x=11. At this hole density the $c$-axis resistivity is metallic above
80K, while for smaller densities the system is insulating at all
temperatures below 300K.  The analysis of Osafune et al. (1997) made
possible to isolate the contributions of the ladders from those of the
chains. These authors showed that holes are actually transferred from
chains to ladders upon Ca-substitution, as explained below. This result
is in agreement with the conclusions of several other studies.

Fig.~26 contains the optical conductivity along the $c$-axis obtained by
Osafune et al. (1997) for several Ca-concentrations and even for an
Y-doped sample (x=3). It is clear from the figure that the conductivity
in the low-energy region below 1.2 eV increases with x, while the
charge-transfer spectral weight decreases, suggesting a transfer of
weight similarly as it occurs in the 2D cuprates when an insulator
parent compound is hole-doped (see, for instance, Uchida et al., 1991).
The analogy between results for ladders and planes clearly suggests that
Ca-doping produces a redistribution of holes, moving them from the
chains to the ladders.  If the holes would be populating the chains,
substantially different results would be expected instead of the
observed doping of a charge-transfer parent compound. In addition, the
results found experimentally are similar to those found by numerical
simulations of models for ladders by Hayward, Poilblanc and Scalapino
(1996).  The low-energy integrated optical conductivity up to $\omega$,
namely $\rm N_{eff}(\omega)$, is shown in Fig.27, and it also resembles
results for the 2D cuprates. Based on these analogies, Osafune et al.
(1997) estimated the concentration of holes on the ladders as a function
of Ca-doping. The result is shown in the inset of Fig.27. At x=11, the
nominal valence of Cu is 2.2, which corresponds to 0.2 holes per Cu on
the ladders. Once again, these doping concentrations resemble those
reached in the two-dimensional square-lattice cuprates.

Interesting recent results by Osafune et al. (1999) using optical
techniques have contributed further to the understanding of doped ladder
materials.  These authors concentrated their efforts on the infrared
part of the spectra of single crystals 14-24-41 with x=8 and 11, as
representatives of the low- and high-doping regions of the ladders
(having 0.16 and 0.20 holes per ladder-Cu, respectively, based on the
previous analysis of Osafune et al. (1997)). The resistivity along the
leg ($c$) and rung ($a$) directions is shown in Fig.28. At x=8 both
directions show an insulating-like behavior, while at x=11 there is a
range of temperatures where the system is metallic in the leg-direction
and insulating in the rung-direction.  An interesting feature in the
$a$-axis optical conductivity $\rm \sigma_a(\omega)$ both at x=8 and 11
is shown in Fig.29. A pseudogap in the low-frequency region appears upon
lowering the temperature (see also Ruzicka et al., 1998). It is
reasonable to expect that the insulating properties that exist in
$\rho_a$ vs T are related with this pseudogap, as it occurs in
underdoped high-$\rm T_c$ cuprates when the pseudogap and resistivity in
the interplane direction are contrasted (see Homes et al., 1993). In
addition, the NMR experiments of Magishi et al. (1998) showed that the
temperature $\rm T^*$ at which a gap opens in the doped ladders
coincides with the analogous temperature arising from $\rho_a(T)$ data.

In contrast to the pseudogap in the $a$-direction, the $c$-axis optical
conductivity $\rm \sigma_c(\omega)$ was found by Osafune et al. (1999)
to be dominated by a low-frequency peak which rapidly grows with
lowering temperature (Fig.30). This is not a Drude peak since it is
located at a finite frequency $\sim 100 cm^{-1}$ and $\sim 50 cm^{-1}$
for x=8 and 11, respectively. Osafune et al. (1999) have interpreted
this result as arising from a charge-density-wave CDW mode of hole
pairs.  In addition, independent studies of the electrodynamics of x=0,
5, and 12 14-24-41 samples by Ruzicka et al. (1998) led to a similar
proposal of a CDW of hole pairs at low-temperatures. Pseudogap features
along the $c$- and $a$-axis were also observed by these authors. Also
Owens, Iqbal and Kirven (1996) using microwave loss measurements
detected superconductive pairing fluctuations close to a CDW
instability.  The proposed hole-pair CDW state is certainly possible
since it is well-known from theoretical calculations that
superconductivity and charge-ordering are competing states in many
models (see Dagotto, Riera, and Scalapino, 1992; Noack, White and
Scalapino, 1994; Tsunetsugu, Troyer and Rice, 1995). Indeed the
effective mass that Osafune et al. (1999) obtained from their analysis
is very large suggestive of the collective nature of the
excitation. Thus, the proposed picture is that hole-pairs are formed and
confined on the ladders, as predicted by theory, at small hole
concentration.  As the hole density grows they seem to form a CDW state
of pairs. Upon increasing the hole concentration and working at
intermediate temperatures the CDW may be depinned by thermal
fluctuations and both pairs and single holes could contribute to the
resistivity along the $c$-axis.  Needless to say, it is very important
to extend these results obtained at ambient pressure into the
superconducting regime at high pressure, where both resistivities
indicate a metallic behavior. Finally, note that very recent Cu-NQR and
NMR experiments by Ohsugi et al. (1999) for the Ca(x=11.5) ladder have
revealed three-dimensional magnetic ordering below $\rm T_N = 2.2K$,
with small moments appearing on the ladders and large moments on the
chains. It was suggested that the CDW state of pairs plays an essential
role in stabilizing the magnetic state.


\vspace{0.6cm}
\centerline{\it 3.3.4 Nuclear Magnetic Resonance:}
\vspace{0.6cm}

A comprehensive Cu NMR study of single crystals of $\rm Sr_{14-x} Ca_x
Cu_{24} O_{41}$ has been recently reported by Magishi et
al. (1998). These authors analyzed the $\rm ^{63}Cu$ NMR spectrum,
Knight shift, nuclear spin-lattice relaxation rate $\rm 1/T_1$, and the
spin-echo decay rate $\rm 1/T_{2G}$ for samples with $\rm x=0$, 6, 9,
and 11.5.  The main reported results emerging from this study are the
following:

i) Through the Knight shift and $\rm 1/T_1$ measurements, the spin-gap
was inferred. The analysis of both quantities suggest that the gap
decreases as the Ca-concentration grows (see Fig.31). Qualitatively
similar results were found using polycrystals by Tsuji et al. (1996).

ii) In particular, for x=11.5 and ambient pressure the gap was found to
 be finite and approximately 300K (actually from Knight shift the result
 is 270K, while from $\rm 1/T_1$ it is 350K). This is important since at
 this Ca-concentration superconductivity has been observed upon the
 application of pressure by Nagata et al. (1997). Fig.32 shows the
 overall phase diagram obtained using NMR techniques. Note that the
 spin-gap seems to flatten as x continues growing, and it becomes
 unclear at what Ca-composition the gap will finally vanish if at
 all. The flattening of the spin-gap with Ca-composition is actually in
 contradiction with the previous results by Kumagai et al. (1997) using
 polycrystals.  Note that in spite of the presence of a nearly constant
 spin-gap at large x, the conductivity changes substantially with
 increasing Ca-concentration.

iii) From data corresponding to $\rm T_{2G}$, the spin correlation
length $\xi$ was obtained experimentally. It was found to decrease as x
grows, as expected. Assuming this length is dominated by the mean
distance among mobile holes, at Ca11.5 a hole concentration of x=0.25
holes per $\rm Cu_2 O_3$ was obtained.  This value is in good agreement
with the hole concentration x=0.22 inferred from optical measurements by
Osafune et al. (1997).

iv) At a temperature of around 60K, charge localization takes place.
The resistivity increases in this regime.

v) The resistivity has a very interesting behavior with temperature as
already discussed elsewhere in this review.  At ``high'' temperatures
the resistivity along the c-axis (i.e. along the legs of the ladders) is
linear, while in the perpendicular direction the resistivity is
T-independent and larger by an order of magnitude. The spin dynamics was
found to be consistent with the S=1/2 1D Heisenberg model.  In an
intermediate temperature regime the resistivity deviates slightly from
the linear T-dependence, again similarly as it occurs in the 2D doped
cuprates in the pseudogap region, while $\rho_a$ grows as T
decreases. These results suggest that in this regime pairs are expected
to be formed, but they are confined into the ladders.  In the ``low''
temperature regime, charge localization takes place presumably through
the localization of the hole pairs on the ladders.  In order to
understand whether the application of pressure deconfines the hole pairs
from the ladders, NMR studies should be carried out in the regime of
high pressure and some results are already available as discussed below.

Several other NMR experiments on ladder materials have also been
reported. They include Cu-NQR measurements for $\rm Sr_{14} Cu_{24}
O_{41}$ by Carretta, Vietkin, and Revcolevschi (1998), and $\rm ^{17}O$
NMR in doped and undoped $\rm A_{14} Cu_{24} O_{41}$ ($\rm A_{14} = La_6
Ca_8, Sr_{11} Ca_{3}$) by Imai et al. (1998).  A comprehensive $\rm
^{63,65}Cu$ NMR/NQR study by Takigawa et al. (1998) on single crystals
of undoped 14-24-41 reported two distinct resonance spectra for the
chains, assigned to magnetic Cu-sites and nonmagnetic Zhang-Rice
singlets. Microwave frequencies studies of the dynamical magnetic
susceptibility by Zhai et al. (1999) show indications of charge ordering
in the doped 14-24-41 material.

More recently, Mayaffre et al. (1998) reported $\rm ^{63}Cu$ NMR
measurements of the Knight shift and relaxation time $\rm T_1$ on the
x=12 Ca-doped 14-24-41 compound at high pressure. When the system
becomes superconducting at 5K and 31 kilobars, those authors argued that
the spin-gap collapsed based on, e.g., the relaxation rate $\rm
T^{-1}_1$ vs temperature data shown in Fig.33.  In addition, since they
observe that superconductivity starts when the ladder spin-gap
collapses, they argue that the ladder is responsible for the
superconducting phase.  It may occur that the preformed pairs on the
two-leg ladders become dissociated upon pressure, making possible the
hole hopping in the direction perpendicular to the legs.  Regarding this
interesting NMR data by Mayaffre et al. (1998), this author believes
that it is important to obtain an independent confirmation of such an
important result.  Very recently, Mito et al. (1998) using once again
NMR techniques at high pressure for a single crystal of 14-24-41 with
x=12 studied the Knight shift and $\rm T_1$ vs temperature up to
pressures 1.7~GPa. They found a robust spin gap in their studies of
magnitude quite similar to the result obtained at ambient pressure. It
is difficult to understand how this gap can disappear so abruptly as the
pressure goes up to 3~GPa, as claimed by Mayaffree et al. (1998).
Hopefully Mito et al. will be able to extend their analysis to higher
pressures in the near future.  In addition, it should be investigated to
what extend the collapse of the spin-gap reported by Mayaffre et
al. (1998) corresponds to the creation of a small population of
low-energy excitations by pressure, leaving large weight at the original
spin-gap position which may remain mostly untouched as a high-energy
feature, similarly as it has been shown theoretically to occur in the
t-J model for two-leg ladders doped with vacancies by Martins et
al. (1996,1997).  In other words, it would be interesting to investigate
if the results of Mayaffre et al. (1998) correspond to the replacement
of a hard-gap by a ``pseudogap'', namely a clear depletion in the spin
density of states at low-energy which, however, does not vanish at any
finite energy. Such a behavior is known to occur in the high temperature
superconductors, and it would be very important to analyze to what
extend doped ladders present a similar pseudogap phenomenology.
Actually very recent NMR experiments by Auban-Senzier et al. (1999)
already support the conjecture discussed here that a pseudogap appears
in doped ladders at the high pressures needed to reach the
superconducting state.

\vspace{0.6cm}
\centerline{\it 3.3.5 Inelastic Neutron Scattering:}
\vspace{0.6cm}

Inelastic neutron scattering (INS) experiments on the 14-24-41
compound have also been reported. Using this technique, Eccleston, Azuma
and Takano (1996) analyzed a polycrystaline sample of x=2.8 14-24-41
and they observed a gap of 35 meV for the ladder. Broad features at 10
meV corresponding to chain excitations were also reported.  More
recently, Eccleston et al. (1998) presented the first INS study which
has been able to access the spin ladder excitations using single
crystals of a cuprate ladder material. A ladder spin-gap of $32.5 \pm
0.1$ meV was reported with the scattering function $\rm S({\bf
q},\omega)$ sharply peaked about the antiferromagnetic zone center. A
band maximum at $193.5 \pm 2.4$ meV was also observed. Excitations at
energy transfers of 15 meV or below were assigned to the chains present
in 14-24-41. INS results with emphasis on the physics of chains have
also been reported recently by Regnault et al. (1998) and Matsuda et
al. (1998), adding to the studies of Hiroi et al. (1996), Matsuda et
al. (1996), Matsuda et al. (1997), D. E. Cox et al. (1998), and others
that have concentrated their efforts on the analysis of that portion of
the structure of 14-24-41. The rich physics of dimerized chains
observed in this compound will not be reviewed here since the focus of
the paper is on the physics of ladders, but it is clear that its
analysis deserves further attention.

Recent inelastic neutron scattering results by Katano et al. (1999) for
single crystal of x=11.5 14-24-41 at both ambient pressure and high
pressure of 0.72, 2.1 and 3.0 GPa produced surprising
conclusions. Contrary to the NMR results of Magishi et al. (1998),
Katano et al. (1999) reported a spin-gap in the doped system which does
$not$ change from that of the parent material. This occurs even under
high pressure. The origin of these differences between NMR and INS is
unclear at present (see Fig.34).  Note that for x=11.5 in the 14-24-41
material superconductivity appears between 3.5 GPa and 8 GPa, i.e. very
close in pressure to where the last set of experiments by Katano et
al. (1999) have been carried out.  Note also that the INS data, if valid
at slightly larger densities, would be in contradiction with the results
of Mayaffre et al. (1998) that reported the absence of a spin-gap in the
x=12 14-24-41 compound under 3.2 GPa of pressure.  This author would
like to remark that the analysis of the pressure dependence of the INS
data into the pressure range of the superconducting state is very
important, it should certainly continue, and further work hopefully will
solve the discrepancy between INS and NMR results.

\vspace{0.6cm}
\centerline{\it 3.3.6 Photoemission and Angle-Resolved Photoemission}
\centerline{\it $\rm A({\bf p}, \omega)$:} 
\vspace{0.6cm}

Angle-resolved photoemission studies of the ladder compound $\rm
Sr_{14-x} Ca_x Cu_{24} O_{41}$ have been carried out by Takahashi et al.
(1997) and Sato et al. (1997) using single crystals with x=0 and 9.
These authors found two dispersive features in their data.  One has an
energy dispersion of about 0.5 eV, matching well the periodicity of the
ladder. The minimum binding energy is located at momentum $\rm ka=\pi/2$
for x=0, in rough agreement with numerical studies of the t-J model by
Tsunetsugu, Troyer and Rice (1994,1995); Troyer, Tsunetsugu, and Rice
(1996); Haas and Dagotto (1996); and Martins, Gazza, and Dagotto (1999),
that locate this minimum close to, although not exactly at, $\pi/2$ in
the bonding band of isolated ladders.  The other features in the
photoemission results by Takahashi et al.  (1997) are deeper in energy,
and they seem to correspond to chain excitations.

Other studies of the electronic structure of
(La, Sr, Ca)$_{14}$Cu$_{24}$O$_{41}$ using core-level photoemission
spectroscopy have been presented by Mizokawa et al. (1998), and the
results compared against another ladder compound, $\rm La_{1-x} Sr_x Cu
O_{2.5}$, as well as the high-$\rm T_c$ material $\rm La_{2-x} Sr_x Cu
O_4$. These authors observed that the chemical potential shift in doped
14-24-41 is considerably suppressed compared with the other two
compounds, as shown in Fig.35.  Mizokawa et al. (1998) interpreted this
as an indication of a non-Fermi liquid state on the 14-24-41 ladders
being more robust against hole doping than in other copper oxides.

This interesting result shows that much remains to be learned about
ladders using photoemission experiments and hopefully in the future
information as rich as reported for the high-$\rm T_c$ compounds will be
also available for ladders.  In this context theory is still ahead of
experiments and a considerable amount of information is already
available to contrast experiments with results for the t-J model (see
for instance Riera, Poilblanc and Dagotto, 1999). In particular, Martins
et al. (1999) using a novel technique that allowed them to study
clusters containing $2 \times 20$ sites recently reported the evolution
with doping of the dominant peak in the one-particle spectral function
$\rm A({\bf q}, \omega)$, presumed to be a quasiparticle, as a function
of density. The result is shown in Fig.36 for the bonding band $\rm q_y
= 0$.  At x=0.0 (undoped limit) the bonding band is very narrow. At
x=0.125, it remains flat near $(\pi,0)$ quite similarly as in
experiments for the 2D cuprates. A gap develops at this density due to
pair formation. As x keeps on growing the photoemission and
inverse-photoemission bands loose their flatness, and start deforming
resembling more a quasi-non-interacting dispersion which is reached at
x=0.5.  This theoretical result is the first one available where the
evolution of the one-particle spectral function with doping has been
reported between a limit dominated by short-range antiferromagnetic
correlations and a quasi-non-interacting regime. The analog result for
the 2D cuprates is still unknown.

\vspace{0.6cm}
\centerline{\bf 3.4 The Strong-Coupling Ladder Compound}
\centerline{\bf Cu$_2$(C$_5$H$_{12}$N$_2$)$_2$Cl$_4$ }
\vspace{0.6cm}

Spin ladders belonging to the organic family of materials have also been
synthesized. In particular, the compound $\rm Cu_2(C_5 H_{12} N_2 )_2
Cl_4$ has Cu-ions which are coupled antiferromagnetically in isolated
ladders, as remarked by Chaboussant et al. (1997a,1997b).  The schematic
structure of this compound is shown in Fig.37.  The coupling between
coppers is mediated by chlorine ions (not shown in Fig.37).  The
orbitals involved in the process have a small overlap producing a
reduced value for that coupling. Experimentally, the Heisenberg
exchanges were found to be $\rm J_{\perp} = 13.2K$ and $\rm J_{||} =
2.4K$, following the notation of Fig.37, leading to an anisotropic
ladder that belongs to the strong coupling limit since $\rm
J_{\perp}/J_{||} = 5.5$ (Chaboussant et al., 1997b).  In this case, the
rung spin-singlet must be very strong, leading to a regime which is
under good control in the theoretical calculations using t-J-like
models. The physics in this strong coupling region of parameter space is
expected to be smoothly connected to the isotropic regime $\rm J_{\perp}
\sim J_{||}$, as discussed in Sec.2 and by Dagotto, Riera and Scalapino
(1992).  In addition, the intrinsic small values of the Heisenberg
couplings allows for investigations in static magnetic fields of
comparable strength, providing a window of research that is not possible
in the inorganic cuprate ladders with exchange couplings of order $\rm
0.1eV \sim 1000K$.  Since at large $\rm J_{\perp}/J_{||}$ the dominant
spin configurations are singlets along the rungs, the lowest-energy
excitations correspond to local rung-triplets which can move along the
ladder. A finite density of these excitations is needed to represent the
physics of ladders in high magnetic fields of strength large enough to
produce a finite magnetization.

After NMR techniques were applied to the organic ladder compound by
Chaboussant et al. (1997b), a spin gap $\rm \Delta \sim J_{\perp} -
J_{||}$ was reported, in agreement with theoretical expectations at
large $\rm J_{\perp}/J_{||}$.  In addition, the gap obtained from the
static susceptibility $\chi$ and the dynamical spin structure factor
obtained from the NMR measurements were found to be identical, as
predicted by theory.  Further work by Chaboussant et al. (1998a), still
using NMR methods, analyzed the transition from the gapped spin-liquid
state to a gapless magnetic regime. At the critical magnetic field $\rm
H_{c1} \sim 7.5T$ where the gap closes, the system is in a quantum
critical regime where the temperature is the only relevant scale. At a
larger field, $\rm H_{c2}$, the system becomes fully
polarized. Measurements of the susceptibility and magnetization in this
regime are in good agreement with calculations by Weihong, Singh, and
Oitmaa (1997) (see also Hayward, Poilblanc and Levy (1996)).  More
recent studies of the organic ladder by Calemczuk et al. (1999), based
on specific heat data, reported a possible incommensurate gapped state
due to a small magneto-elastic coupling of the isolated ladders with the
3D lattice.  Associating the rung-singlet and rung-triplet states, the
latter having spin projection along the magnetic field, with fictitious
``spin'' states up and down, Chaboussant et al. (1998b) argued that the
physics of strongly coupled ladders in high magnetic fields can be
mapped into a 1D XXZ Heisenberg spin model which is known from previous
investigations to present incommensurate phases in magnetic
fields. Calculating the magnetization with the Bethe Ansatz applied to
the XXZ model and comparing the results with the experimental
magnetization leads to an excellent agreement (see Fig.38). Since the
XXZ model is equivalent to spinless fermions which in 1D form Luttinger
liquid (LL) states, a regime with these characteristics was conjectured
to exist between $\rm H_{c1}$ and $\rm H_{c2}$, as indicated in
Fig.39. At very low temperatures the small couplings between ladders
will destabilize the LL leading to a charge ordered state, as also
sketched in Fig.39.

Other recent developments in this context include the presence of
plateauxs in the magnetization curve in the vicinity of $\rm (H_{c1} +
H_{c2})/2$, as argued by Cabra, Honecket and Pujol (1997), Cabra and
Grynberg (1998) and others.  Calculations for coupled ladders in a
magnetic field have also been recently reported by Giamarchi and Tsvelik
(1998). Other studies in magnetic fields were presented by Schmeltzer
and Bishop (1998).  In addition, some of the results derived in terms of
a ladder with only two couplings $\rm (J_{\perp},J_{||})$ are
controversial since Hammar et al. (1998) argued that extra ladder
couplings, as well as multimagnon excitations and interladder couplings,
may be needed to properly reproduce their inelastic neutron scattering
and specific heat data (see also Hayward, Poilblanc, and Levy (1996)).
Overall it is clear that the analysis of ladder systems under high
magnetic fields have already revealed an unexpected rich structure, and
theoretical and experimental efforts should continue exploring this
interesting regime in parameter space.

Finally, doping with impurities the organic ladder can reveal
interesting physics.  Using computational techniques, Laukamp et
al. (1998) predicted that Zn-doping of the organic strong-coupling
ladder should not produce important changes in the spin-gap, contrary to
what is observed in other Zn-doped ladders.  Experiments are already in
agreement with this prediction (see Deguchi et al. (1998)).

\vspace{0.6cm}
\centerline{\bf 3.5 Vanadium-based Ladder Compound CaV$_2$O$_5$ } 
\vspace{0.6cm}

Interesting studies by Iwase et al. (1996) showed that the vanadate $\rm
Ca V_2 O_5$ has a large spin-gap of about 500K.  The structure of this
compound corresponds to a depleted two-dimensional square lattice. It
consists of a so-called trellis-lattice with two-leg ladders containing
V-ions in a $\rm V^{4+}$ state of spin-1/2, with the idealized structure
shown in Fig.40. The ladders are coupled through $90^o$ oxygen bonds
which are expected to produce a weak exchange, geometry similar as in
some of the cuprate ladder materials, although involving the vanadium
$\rm d_{xy}$-orbitals.  Actually, studies by Normand et al. (1997)
reported couplings $\rm J_{\perp}/J_{||} \sim 4.3$ and $\rm J_{||}/J'
\sim 3.5$, in the notation of Fig.40.  However, recent Quantum Monte
Carlo simulations of the trellis-lattice Heisenberg model by Miyahara et
al. (1998) produced a magnetic susceptibility vs temperature in
agreement with experiments if $\rm J_{\perp} \sim 670K$, $\rm J_{||}
\sim 67K$, and $\rm J' \sim 45K$, namely using $\rm J_{||} \sim J'$.
This result suggests that the dominant structures in the vanadate
ladders are the dimers or rungs with spins coupled by $\rm J_{\perp}$,
rather than the ladders themselves.  The resolution of this discrepancy
is important to decide whether $\rm Ca V_2 O_5$ should be considered as
a two-leg ladder material.

More recently, the compound $\rm Mg V_2 O_5$ has also been studied (see
Millet et al., 1998). The V-oxide planes of this material are very
similar to those of $\rm Ca V_2 O_5$. However, the observed gap is just
15K, much smaller than the 500K reported for the latter. The difference
was explained by Millet et al. (1998) on the basis of a frustrated
coupled ladder model.  Other recent developments involve the
spin-Peierls compound $\rm Na V_2 O_5$. Based on the determination of
the crystal structure of this compound by x-ray diffraction and on
density-functional calculations, Smolinski et al. (1998) argued that
$\rm Na V_2 O_5$ can be considered a quarter-filled two-leg ladder
system.  The exchange interaction of $\rm Na V_2 O_5$ across the rungs
was found to be larger than along the legs by a factor 5.  Since this
compound and $\rm Ca V_2 O_5$ are isostructural, it was also argued that
the latter should also be regarded as a (half-filled) ladder system
instead of only having isolated dimers (see discussion in previous
paragraph).  Very recent results by Lemmens et al. (1998) using Raman
scattering applied to $\rm Na V_2 O_5$ showed an interesting sequence of
magnetic bound states at low temperatures.  The importance of doping
$\rm Na V_2 O_5$ away from quarter-filling was remarked by Smolinski et
al. (1998).  Clearly this system provides a new playground to test the
interplay between spin, charge, orbital and lattice degrees of freedom
in a ladder geometry, and further work should be pursued in this
context.


\vspace{0.6cm}
\centerline{\bf 3.6 The Ladder Compound KCuCl$_3$ } 
\vspace{0.6cm}

Recently, Tanaka et al. (1996) have presented interesting arguments
suggesting that the material $\rm K Cu Cl_3$ can also be described as a
double spin-chain system.  The crystal structure of this compound is
shown in Fig.41. The main feature is the double chain of edge-sharing
$\rm Cu Cl_6$ octahedra along the $a$-axis. Also in Fig.41 it is shown
the expected pattern of Heisenberg exchange couplings arising from such
a structure. They involve not only the standard rung- and leg-couplings,
but also a plaquette-diagonal coupling. Their values are expected to be
small since $\sim 90^o$ bonds are involved in the structure.  To
investigate the nature of the ground state of this material, Tanaka et
al. (1996) studied the temperature dependence of the magnetic
susceptibility.  The low-temperature results can be fit very well with
the formula $\rm \chi \sim (1/\sqrt{T}) exp(-\Delta/k_BT)$ proposed by
Troyer, Tsunetsugu and W\"urtz (1994). The gap obtained by this
procedure was found to be 35K.  Large scale quantum Monte Carlo
calculations performed by Nakamura and Okamoto (1998) on the spin model
believed to describe this material, found an agreement with the
experimental data within 1\% relative errors in the full range of
temperatures investigated. The comparison theory-experiment involved
neutrons scattering data obtained by Kato et al. (1998).  The couplings
that reproduce the experimental data the best are $\rm J_1 = 12.3K$,
$\rm J_2 = 4J_1$, and $\rm J_3\sim 0$ suggesting that the extra coupling
$\rm J_3$ may not be necessary.  In addition, note that related
double-spin chain systems, such as $\rm NH_4 Cu Cl_3$, have also been
studied using ESR and other techniques by Takatsu et al. (1997) and
Kurniawan et al. (1999).

The values of the exchange couplings that have been reported for $\rm K
Cu Cl_3$ locate the material in the strong rung-coupling limit,
similarly as it occurs for the $\rm Cu_2(C_5 H_{12} N_2 )_2 Cl_4$
compound, as explained before in this review. For this reason it is
natural for this author to suggest the use of (probably pulsed) high
magnetic fields for the study of $\rm H_{c1}$, which is the magnetic
field at which the gapped spin state becomes gapless (for results with
high fields see Shiramura et al. (1997)).  The motivation is to
investigate whether the interesting physics reported by Chaboussant et
al. (1998b) is also present in $\rm K Cu Cl_3$, especially regarding the
possible presence of incommensurate spin correlations.  Also the
chemical replacement of Cu-ions by Zn- or Ni-ions would add to the
ongoing discussion on the influence of vacancies on the physics of
ladders and spin-Peierls systems, as discussed earlier in the paper (see
Martins et al., 1997, and references therein. See also Sandvik, Dagotto
and Scalapino, 1997).

\section{Conclusions} 

In this review the experimental aspects of the study of ladder materials
have been discussed, with emphasis on the relation with the high-$\rm
T_c$ cuprates and contrasting the results with theoretical
predictions. After a considerable effort, a variety of experiments have
clearly confirmed the predicted existence of a spin-gap in two-leg
ladders, a feature that in the form of a pseudogap also appears in the
underdoped 2D cuprates. This is an important achievement of theoretical
calculations, showing that at least in spin systems and low-dimensions
the models and many-body tools available are powerful enough to make
reliable predictions.  Even in the context of Zn-doped ladders a good
agreement theory-experiments has been reported. Here small amounts of
vacancies produce a considerable enhancement of the antiferromagnetic
correlations in their vicinity and the ladders develop long-range order
out of an apparently disordering procedure.  Some other undoped ladders
are in the strong rung-coupling regime and the values of their couplings
are such that they can be study in high magnetic fields, leading to a
rich phase diagram that is also in good agreement with theoretical
studies. Overall, the analysis of ladders in the absence of mobile holes
seems under good control.

Once carriers are added to some of the available ladder compounds, a
metal-insulator transition has been observed. At ambient pressure the
carriers seem confined to move along the ladders. As the pressure
increases, the hole movement along the rung direction is enhanced and
superconductivity appears in the 14-24-41 ladder. Pressure is presumed
to increase the interladder couplings.  Actually it has become clear
that such couplings are not negligible in several real ladders even in
the undoped regime, causing transitions to antiferromagnetic states at
low temperatures.  While most authors agree with the notion that hole
pairs are created on the ladders, probably by the mechanism predicted by
the theorists, it is still under discussion what triggers the transition
to the superconducting phase. It may occur that the interladder hopping
is important to prevent the system from forming a charge-density-wave of
pairs, that some experiments have detected at low-pressure. In this
respect a large pressure is needed to provide extra mobility to the
carriers, but too much pressure would conspire against the formation of
pairs on the individual ladders. This may explain the existence of an
optimal pressure. In addition, theoretical studies have shown that in
the undoped regime an apparently small interladder coupling may destroy
the spin-gap and transform the ground state into a gapless
antiferromagnet. Since the presence of spin-gap features is important
for the superconductivity predicted by the theorists, once again it is
concluded that too much pressure may be detrimental for
superconductivity.

Some experimental aspects are still controversial, notably the value of
the spin-gap as the hole concentration and pressure changes in
14-24-41. Some NMR experiments associate the closing of the spin-gap
with the appearance of superconductivity. However, other NMR and neutron
scattering experiments at slightly smaller pressures report a very
robust spin-gap. Hopefully these differences will be clarified in the
near future.  Resistivity measurements for this compound have found that
the system becomes metallic both along the leg and rung directions at
the optimal pressure. Similarities of this behavior with results for
two-dimensional cuprates have been here remarked by the author. A
possible underdoped-like regime could exist in 14-24-41 in a narrow
pressure window.  Note that regardless of the 1D vs 2D character of the
superconductivity, x-ray experiments have confirmed the survival of the
ladder substructures in the 14-24-41 material under high
pressure. Then, this compound is the $first$ copper-oxide superconductor
with a non-square-lattice arrangement. In addition, the resistivity
versus temperature on the 14-24-41 ladder is linear at several
concentrations adding to the list of common aspects with the 2D
cuprates. Several other similarities have been discussed in the paper.
Certainly much can be learned out of a superconducting compound that has
the same Cu and O building blocks as the high-$\rm T_c$ materials.

The overall conclusion is that the analysis of ladder compounds has
already developed into a fruitful area of research, and quite
challenging and interesting results have become available in recent
years. Among the most notable discoveries are a novel copper-oxide
superconducting material with a ladder-dominated structure, and several
others compounds with a clear spin-gap in their spectrum. Further work
in the context of ladders will likely keep on providing key information
to unveil the mystery of the origin of superconductivity in the cuprates
and the behavior of electrons in transition-metal-oxides in general.

\section{Acknowledgments}

Part of the results discussed here have emerged from conversations and
discussions with several colleagues working on ladder compounds.  The
author especially thanks J. Akimitsu, M. Azuma, T. Barnes, G.
Chaboussant, L. Degiorgi, A. Fujimori, Z. Hiroi, D. J\'erome,
D. C. Johnston, M-H. Julien, Y. Kitaoka, A. Moreo, D. Poilblanc,
T. M. Rice, J. Riera, D. J. Scalapino, M. Takano, S. Uchida, for their
comments.  The author is supported in part by grant NSF-DMR-9814350.

\pagebreak

\section*{FIGURE CAPTIONS}

\begin{enumerate}

\item{(a) Schematic representation of the ground state of a
two-leg ladder. Pairs of spins along the same rung tend to form a spin
singlet; (b) Individual holes added to the ladders destroy spin
singlets; (c) To minimize the energy damage caused by the addition of
holes, these holes tend to share the same rung forming hole
bound-states. Figure reproduced from Hiroi (1996).}

\item{ Spin-gap $\rm \Delta_{spin}$ versus the ratio $\rm J'/J$
for the two-leg ladder, with the convention that $\rm J'$ is the
coupling along the rungs and $\rm J$ along the legs. The results are
extrapolations to the bulk limit of data obtained on finite $\rm 2
\times N$ clusters.  The figure is reproduced from Barnes et al. (1993).}

\item{Susceptibility as a function of temperature for the
single chain and the Heisenberg ladders with up to six legs, obtained by
Frischmuth, Ammon and Troyer (1996) using a Monte Carlo algorithm. The
upper and lower parts of the figure differ in the temperature range
analyzed.  At low-temperature, frame (b) clearly shows that the odd- and
even-leg ladders have a qualitatively different behavior, as described
in the text.}

\item{Log-log plot of various correlation functions versus the
real space separation $\rm |i-j|$ for a $2 \times 30$ open chain with
density $\rm \langle n \rangle = 0.8$ and $\rm J/t=1.0$. The dashed line
has a slope -2 and the dotted line -1. Figure taken from Hayward et
al. (1995), where more details can be found on the definition of the
correlation functions used here.}

\item{Dynamical spin structure factor $\rm S({\bf q}, \omega)$
for the $\rm q_y=\pi$ branch of a $2 \times 16$ cluster with two
holes. The physical meaning of branches (I) and (II) is explained in the
text.  The figure is taken from Dagotto et al. (1998) using a technique
involving a small fraction of the Hilbert space, after the t-J model is
rewritten (exactly) in the rung-basis.}

\item{Schematic representation of (a) the $\rm Cu_2 O_3$
sheets of $\rm Sr Cu_2 O_3$ (from Azuma et al., 1994). The three-leg
ladder $\rm Sr_2 Cu_3 O_5$ is also shown in (b). The filled circles are
$\rm Cu^{2+}$ ions, and $\rm O^{2-}$ ions are located at the corners of
the squares drawn with solid lines.}

\item{Temperature dependence of the magnetic susceptibility of
$\rm Sr Cu_2 O_3$ (from Azuma et al., 1994). Details can be found in the
text.}

\item{Similar as Fig.~7 but for the three-leg compound $\rm
Sr_2 Cu_3 O_5$ (from Azuma et al., 1994).  In this case a large
susceptibility is observed even at low temperatures, indicative of the
absence of a spin-gap, in agreement with theoretical expectations.}

\item{Magnetic susceptibility vs temperature for $\rm Sr Cu_2
O_3$ (filled circles). Shown as open symbols are fits to these data
using the calculations of Barnes and Riera (1994) for the isolated
spin-1/2 two-leg ladder at several ratios of the rung Heisenberg
coupling, here denoted by $\rm J'$, and the leg coupling J. Figure
reproduced from Johnston (1996). A comprehensive experimental paper with
updated information about these issues is in preparation (D. Johnston,
private communication).}

\item{Magnetic susceptibility of $\rm Sr(Cu_{1-x} Zn_x)_2
O_3$ vs. temperature at several Zn-concentrations. The inset shows the
temperature dependence of the inverse susceptibility. The results are
reproduced from Azuma et al. (1997).}

\item{Local staggered susceptibility $\chi_r (-1)^r$ along
the upper leg of a $2 \times 50$ ladder obtained with computational
techniques (from Laukamp et al., 1998). Two vacancies are included in
the system, located at rung 8 on the upper leg and 43 on the lower
leg. Results for three values of the ratio $\rm J_{rung} = J_{\perp}/J$
are shown. For details see Laukamp et al. (1998).}

\item{Structure of $\rm La Cu O_{2.5}$ along the
$c$-axis. Large, medium, and small balls represent La, Cu, and O atoms,
respectively. The figure is reproduced from Hiroi (1996).}

\item{Temperature dependence of the magnetic susceptibility
for $\rm La Cu O_{2.5}$, from Hiroi (1996). The open circles are the raw
data. The solid line through them is a fit including a temperature
independent term, a Curie term due to impurities, plus the bulk
susceptibility expected from a 2-leg Heisenberg ladder. The dotted line
is the susceptibility after subtracting the Curie component. The other
solid line corresponds to $\rm Sr Cu_2 O_3$, for comparison.}

\item{Resistivity of $\rm La_{1-x} Sr_x Cu O_{2.5}$ vs.
temperature for several Sr-compositions (reproduced from Hiroi, 1996).}

\item{Resistivity at 300K of La$_{1-x}$Sr$_x$CuO$_{2.5}$ and 
La$_{2-x}$Sr$_x$CuO$_4$ , and its doping dependence (reproduced
from Hiroi, 1996).}

\item{Doping dependence of the magnetic susceptibility at
several hole concentrations (from Hiroi, 1996).}

\item{(a) Crystal structure of $\rm (Sr,Ca)_{14} Cu_{24}
O_{41}$. (b) and (c) show the $\rm Cu O_2$ chain and $\rm Cu_2 O_3$
ladder subunit, respectively.  Figure reproduced from Magishi et
al. (1998).}

\item{Electrical resistivity $\rho$ of $\rm Sr_{0.4}
Ca_{13.6} Cu_{24} O_{41.84}$ vs temperature below 50K, and under
pressure of 3, 4.5 and 6~GPa, showing the existence of
superconductivity.  The figure is reproduced from Uehara et al. (1996).}

\item{Temperature dependence of the electrical resistivities
at various pressures, for $\rm Sr_{0.4} Ca_{13.6} Cu_{24} O_{41 +
\delta}$. Data reproduced from Isobe et al. (1998).}

\item{Temperature dependence of $\rm T_c$ (onset) as a
function of pressure for $\rm Sr_{0.4} Ca_{13.6} Cu_{24} O_{41 +
\delta}$, reproduced from Isobe et al. (1998).}

\item{Pressure variation of the lattice constants of $\rm
Sr_{0.4} Ca_{13.6} Cu_{24} O_{41 + \delta}$ at low temperature,
reproduced from Isobe et al. (1998). The $c$-direction runs along the
legs of the ladders, $a$ along its rungs, and $b$ is perpendicular to
both.}

\item{Effect of pressure on the temperature dependence of the
resistivity (a) along the leg-ladder direction ($\rho_c$) and (b) across
the rung-ladder direction ($\rho_a$) of single crystal $\rm Sr_{2.5}
Ca_{11.5} Cu_{24} O_{41}$ at the indicated pressures. The inset shows
the pressure dependence of $\rm T_c$. Figure reproduced from Nagata et
al.  (1998a).}

\item{Effect of pressure on the temperature dependence of the
anisotropic ratio $\rho_a/\rho_c$ of single crystal $\rm Sr_{2.5}
Ca_{11.5} Cu_{24} O_{41}$ at the indicated pressures. The inset shows
the same anisotropic ratio but at pressures above 4.5 GPa.  Figure
reproduced from Nagata et al. (1998a).}

\item{In-plane ($\rho_a$) and out-of-plane ($\rho_c$)
resistivity of $\rm Y Ba_2 Cu_3 O_{7-y}$ plotted as a function of
temperature, reproduced from Takenaka et al. (1994). Results are shown
for various oxygen concentrations $\rm 7-y \sim 6.68, 6.78, 6.88$, and
6.93.}

\item{Anisotropic resistivity ratio $\rm \rho_c/\rho_a$
plotted as a function of temperature for various oxygen contents (from
Takenaka et al., 1994).}

\item{$c$-axis optical conductivity $~\sigma_c(\omega)$ of
$\rm Sr_{14-x} A_x Cu_{24} O_{41}$ for various compositions. The figure
is reproduced from Osafune et al. (1997). The anisotropic spectra with
$a$- and $c$-axis polarization are shown for $\rm Sr_3 Ca_{11} Cu_{24}
O_{41}$ in the inset.}

\item{Effective electron number $\rm N_{eff}$ per Cu
(left-hand scale) presented as a function of energy for various
compositions. The valences of both chain- and ladder-Cu estimated for
each composition are plotted in the inset.  Details can be found in
Osafune et al. (1997).}

\item{T-dependence of the resistivities along the legs
($\rho_c$) and rungs ($\rho_a$) of the ladders in the compound $\rm
Sr_{14-x} Ca_x Cu_{24} O_{41}$, with x=8 and 11 (from Osafune et al,
1999).  The temperatures $\rm T^*$ and $\rm T_0$ where $d\rho/dT$
changes sign are indicated for x=11. The anisotropic resistivity
$\rho_a/\rho_c$ vs temperature is plotted in the inset.}

\item{$a$-axis optical conductivity spectra
$\sigma_a(\omega)$ below 2500 cm$^{-1}$ at various temperatures are
shown for x=8 (upper panel) and 11 (lower panel) (from Osafune et al.,
1999). In the inset are shown the spectra below 800 cm$^{-1}$ to show
the redistribution of the oscillator strength of the phonons.}

\item{$c$-axis optical conductivity spectra
$\sigma_c(\omega)$ at various temperatures for x=8 (upper panel) and 11
(lower panel). The optical reflectivity spectra in the low frequency
region are shown in each inset. This figure is from Osafune et al.
(1999), which should be consulted for further details.}

\item{$\rm (1/T_1)_b$ for a magnetic field parallel to the
$b$-axis vs 1/T. Results are shown for several Ca-compositions of $\rm
Sr_{14-x} Ca_x Cu_{24} O_{41}$. The solid lines are fits to $\rm 1/T_1
\sim exp(-\Delta/T)$. The inset contains the relaxation function of the
nuclear magnetization at T=110K in Ca11.5.  The figure is reproduced
from Magishi et al. (1998).}

\item{Spin gap vs the amount of Ca-substitution in $\rm
Sr_{14-x} Ca_x Cu_{24} O_{41}$, reproduced from Magishi et
al. (1998). Also shown is $\rm T_c$ vs pressure at x=11.5. The spin gaps
$\rm \Delta_K$ and $\rm \Delta_{T_1}$ are obtained from the Knight shift
and $\rm T_1$, respectively. $\rm T_L$ is the temperature associated
with the localization of pairs.  For more details see Magishi et
al. (1998).}

\item{Temperature dependence of the $\rm ^{63}Cu$ ladder
relaxation rate $\rm T^{-1}_1$ at 1 bar and 32.2 kbar. Figure reproduced
from Mayaffre et al. (1998).}

\item{Spin-gap as a function of Ca-content in [14,24,41] at
ambient pressure.  The circles show the results from neutron scattering
experiments (Katano et al. (1999)).  The triangles are the results from
the Knight shift in the NMR experiments of Magishi et al. (1998). The
figures is reproduced from Katano et al. (1999).}

\item{Chemical potential shift as a function of hole
concentration of the Cu-O ladder for $\rm (La,Sr,Ca)_{14} Cu_{24}
O_{41}$ compared with that for $\rm (La,Sr)CuO_{2.5}$. These results are
reproduced from Mizokawa et al. (1998). The hole concentration of the
ladder has been taken from the optical study of Osafune et al. (1997).}

\item{Dominant peak in the one-particle spectral function
$\rm A({\bf p},\omega)$ obtained numerically using the t-J model,
parametric with hole density x. The coupling used was J/t=0.4 and the
lattice had up to $2\times 20$ sites. The size of the points is
proportional to the intensity of the peak. Open circles correspond to
photoemission, while full circles are in the inverse-photoemission
region. Figure taken from Martins et al. (1999), where more details
about the calculation can be found.}

\item{Schematic structure of $\rm Cu_2(C_5 H_{12} N_2 )_2
Cl_4$.  The labeled protons contribute to the NMR lines study, as
discussed by Chaboussant et al. (1997b).}

\item{Magnetization of $\rm Cu_2(C_5 H_{12} N_2 )_2 Cl_4$
between 0 and 20 T, at different temperatures, reproduced from
Chaboussant et al.  (1998b). The symbols are experimental data points,
while the solid lines are the fits to the XXZ model (for details, see
the text and Chaboussant et al., 1998b). The critical fields are
indicated.}

\item{Phase diagram of spin-ladders in a magnetic field,
reproduced from Chaboussant et al. (1998b). The quantum critical regime
at finite temperature above $\rm H_{c1}$ is shown. The ``LL'' label
corresponds to a possible Luttinger liquid region. Further details can
be found in Chaboussant et al. (1998b).}

\item{Schematic representation of the trellis-lattice
corresponding to the structure of the vanadates, reproduced from
Miyahara et al. (1998). The full circles are the positions of the
V-ions. The oxygens in between are not shown. The couplings that are
shown correspond to the Heisenberg exchange constants used to model the
behavior of this material.}

\item{The double chain structure in $\rm K Cu Cl_3$ and the
configuration of the hole orbital $\rm d(x^2 - y^2)$ of the $\rm
Cu^{2+}$ ion are shown in the upper panel.  Filled and open circles are
$\rm Cu^{2+}$ and $\rm Cl^-$ ions, respectively. In the lower panel, the
structure of the exchange interactions expected in the present system is
shown. The figure is reproduced from Tanaka et al. (1996).}

\end{enumerate}

\section*{REFERENCES}

\noindent Ando Y, Boebinger GS, Passner A, Kimura T and Kishio K 1995
Phys. Rev. Lett. {\bf 75} 4662-4665.

\noindent Arai M and Tsunetsugu H 1997 Phys. Rev. {\bf B 56}, R4305.

\noindent Asai Y 1994 Phys. Rev. {\bf B 50} 6519.

\noindent Auban-Senzier P, Mayaffre H, Lefebvre S, Wzietek P, 
J\'erome D, Ammerahl U, Dhalenne G and Revcolevschi A 1999 preprint, to
appear in Proceedings of the International Conference on Synthetic Metals 1998.

\noindent Azuma M, Hiroi Z, Takano M, Ishida K and Kitaoka Y 1994 Phys. Rev.
Lett. {\bf 73}, 3463.

\noindent Azuma M, Fujishiro Y, Takano M, Nohara M and Takagi H 1997 Phys. Rev.
{\bf B 55}, R8658.

\noindent Azzouz M, Chen L, and Moukouri S 1994 Phys. Rev. {\bf B 50} 6233.

\noindent Balakirev FF, Betts JB, Boebinger GS, Motoyama N, Eisaki H and 
Uchida S 1998 cond-mat/9808284.

\noindent Balents L and Fisher MPA 1996a Phys. Rev. {\bf B 53} 12133.

\noindent Balents L and Fisher MPA 1996b cond-mat/9611126.

\noindent Barnes T, Dagotto E, Riera J and Swanson E 1993
Phys. Rev. {\bf B 47}, 3196.

\noindent Barnes T and Riera J 1994 Phys. Rev. {\bf B 50}, 6817.

\noindent Bednorz JG and M\"uller KA 1986 Z. Phys. B {\bf 64} 188.

\noindent Boebinger GS, Ando Y, Passner A, Kimura T, Okuya M, 
Shimoyama J, Kishio K, Tamasaku K, Ichikawa N and Uchida S 
1996 Phys. Rev. Lett. {\bf 77} 5417-5420.

\noindent Cabra D, Honecker A and Pujol P 1997 Phys. Rev. Lett. {\bf 79}, 
5126. 

\noindent Cabra D and Grynberg M, preprint, cond-mat/9810263 and
references therein.

\noindent Calemczuk R, Riera J, Poilblanc D, Boucher JP, Chaboussant G,
Levy L and Piovesana O 1999 Eur. Phys. J. B {\bf 7}, 171.

\noindent Carretta P, Vietkin A and Revcolevschi A 1998 Phys. Rev. {\bf B 57},
R5606.

\noindent Carretta P, Ghigna P and Lascialfari A 1998
Phys. Rev. {\bf B 57}, 11545.

\noindent Carter SA, Batlogg B, Cava RJ, Krajewski JJ, Peck WF Jr
and Rice TM 1996 Phys. Rev. Lett. {\bf 77}, 1378.

\noindent Chaboussant G, Crowell PA, Levy LP, Piovesana O, Madouri A and
Mailly D 1997a Phys. Rev. {\bf B 55}, 3046.

\noindent Chaboussant G, Julien M-H, Fagot-Revurat Y, Levy LP, Berthier C,
Horvatic M and Piovesana O 1997b Phys. Rev. Lett. {\bf 79}, 925.

\noindent Chaboussant G, Fagot-Revurat Y, Julien M-H, Hanson M, Berthier C,
Horvatic M, Levy LP and Piovesana O 1998a Phys. Rev. Lett. {\bf 80}, 2713.

\noindent Chaboussant G, Julien M-H, Fagot-Revurat Y, Hanson M, Levy LP, Berthier C,
Horvatic M and Piovesana O 1998b 
Eur. Phys. J. B {\bf 6}, 167.

\noindent Cox DE, Iglesias T, Hirota K, Shirane G, Matsuda M, Motoyama N,
Eisaki H and Uchida S 1998 Phys. Rev. {\bf B 57}, 10750.

\noindent Dagotto E and Moreo A 1988 Phys. Rev. {\bf B 38}, 5087

\noindent Dagotto E, Riera J and Scalapino DJ 1992 Phys. Rev. {\bf B 45}, 5744.

\noindent Dagotto E 1994 Rev. Mod. Phys. {\bf 66} 763.

\noindent Dagotto E and Rice TM 1996 Science {\bf 271}, 618.

\noindent Dagotto E, Martins G, Riera J, Malvezzi A and Gazza C 1998 Phys. Rev.
{\bf B 58}, 12063.

\noindent Deguchi H, Sumoto S, Takagi S, Mito M, Kawae T, Takeda K,
Nojiri H, Sakon T and Motokawa M 1998 J. Phys. Soc. Jpn. {\bf 67}, 3707.

\noindent Eccleston RS, Barnes T, Brody J and Johnson JW 1994 Phys. Rev.
Lett. {\bf 73}, 2626.

\noindent Eccleston RS, Azuma M and Takano M 1996 Phys. Rev. {\bf B 53}, R14721.

\noindent Eccleston RS, Uehara M, Akimitsu J, Eisaki H, Motoyama N and
Uchida S 1998 Phys. Rev. Lett. {\bf 81} 1702-1705.

\noindent Emery VJ, Kivelson SA and Zachar O 1997 Phys. Rev. {\bf B 56}, 6120,
and references therein.

\noindent Emery VJ, Kivelson SA and Zachar O 1998 cond-mat/9810155.

\noindent Frischmuth B, Ammon B and Troyer M 1996 Phys. Rev. {\bf B 54} R3714.

\noindent Fujiwara N, Yasuoka H, Fujishiro Y, Azuma M and Takano M 1998 Phys.
Rev. Lett. {\bf 80}, 604.

\noindent Fukuzumi Y, Mizuhashi K, Takenaka K and Uchida S 1996 Phys. Rev. Lett.
{\bf 76}, 684.

\noindent Garrett AW, Nagler SE, Tennant DA, Sales BC and Barnes T 1997a
Phys. Rev. Lett. {\bf 79}, 745.

\noindent Garrett AW, Nagler SE, Barnes T and Sales BC 1997b Phys. Rev. 
{\bf B 55}, 3631.

\noindent Gayen S and Bose I 1995 J. Phys. C {\bf 7} 5871.

\noindent Gazza C, Martins G, Riera J and 
Dagotto E 1999 Phys. Rev. {\bf B 59}, R709.

\noindent Giamarchi T and Tsvelik AM 1998 cond-mat/9810219, and references therein.

\noindent Gopalan S, Rice TM and Sigrist M 1994 Phys. Rev. {\bf B 49}, 8901.

\noindent Greven M, Birgeneau RJ and Wiese UJ 1996 Phys. Rev. Lett. {\bf 77} 1865.

\noindent Greven M, Birgeneau R and Wiese UJ 1997 ``Physics News in 1996'',
Supplement to the APS News, May 1997, page 12.

\noindent Haas S and Dagotto E 1996 Phys. Rev. {\bf B 54}, R3718.

\noindent Hammar P, Reich D, Broholm C and Trouw F 1998 Phys. Rev. {\bf B 57},
7846.

\noindent Hansen P, Riera J, Delia A and Dagotto E 1998 Phys. Rev. {\bf B 58} 6258.


\noindent Hayward CA, Poilblanc D, Noack RM, Scalapino DJ and Hanke W 1995
Phys. Rev. Lett. {\bf 75}, 926.

\noindent Hayward CA, Poilblanc D and Scalapino DJ 1996 Phys. Rev. {\bf B 53},
R8863.

\noindent Hayward CA and Poilblanc D 1996 Phys. Rev. {\bf B 53}, 11721.

\noindent Hayward CA, Poilblanc D and Levy LP 1996 Phys. Rev. {\bf B 54},
12649.

\noindent Hiroi Z, Azuma M, Takano M and Bando Y 1991 J. Solid State
Chem. {\bf 95}, 230.

\noindent Hiroi Z and Takano M 1995 Nature {\bf 377}, 41.

\noindent Hiroi Z 1996 J. of Solid State Chem. {\bf 123}, 223.

\noindent Hiroi Z, Amelinckx S, Van Tendeloo G and Kobayashi N 1996 Phys.
Rev. {\bf B 54}, 15849.

\noindent Homes CC, Timusk T, Liang R, Bonn DA and Hardy WN 1993
Phys. Rev. Lett. {\bf 71}, 1645.


\noindent Imai T, Thurber KR, Shen KM, Hunt AW and Chou FC 1998
Phys. Rev. Lett. {\bf 81} 220.

\noindent Ishida K, Kitaoka Y, Asayama K, Azuma M, Hiroi Z and  Takano M 1994
J. Phys. Soc. Jpn. {\bf 63}, 3222.

\noindent Ishida K, Kitaoka Y, Tokunaga Y, Matsumoto S, Asayama K, Azuma M,
Hiroi Z and Takano M 1996 Phys. Rev. {\bf B 53}, 2827.

\noindent Isobe M, Ohta T, Onoda M, Izumi F, Nakano S, Li J, Matsui Y,
Takayama-Muromachi E, Matsumoto T and Hayakawa H 1998 Phys. Rev. {\bf B
57}, 613.

\noindent  Iwase H, Isobe M, Ueda Y and Yasuoka H 1996 J. Phys. Soc.
Jpn. {\bf 65} 2397.

\noindent Johnston DC, Johnson JW, Goshorn DP and Jacobson AP 1987
Phys. Rev. {\bf B 35} 219.

\noindent Johnston DC 1996 Phys. Rev. {\bf B 54}, 13009.

\noindent Kadono R, Okajima H, Yamashita A, Ishii K, Yokoo T, Akimitsu J,
Kobayashi N, Hiroi Z, Takano M and Nagamine K 1996 Phys. Rev. {\bf B
54}, R9628.

\noindent Katano S, Nagata T, Akimitsu J, Nishi M and Kakurai K 1999
Phys. Rev. Lett. {\bf 82}, 636.

\noindent Kato M, Shiota K and Koike Y 1996 Physica C {\bf 258}, 284.

\noindent Kato T, Takatsu K, Tanaka H, Shiramura W, Mori M, Nakajima K
and Kakurai K 1998 J. Phys. Soc. Jpn. {\bf 67}, 752.

\noindent Kimura T, Kuroki K and Aoki H 1996a Phys. Rev. {\bf B 54} R9608.

\noindent Kimura T, Kuroki K and Aoki H 1996b cond-mat/9610200.

\noindent Kishine J and Yonemitsu K 1998 cond-mat/9802185.

\noindent Kivelson SA, Rokhsar DS and Sethna JP 1987 Phys. Rev. {\bf B 35}, 8865,
and references therein.

\noindent Kojima K, Keren A, Luke GM, Nachumi B, Wu WD, Uemura YJ, 
Azuma M and Takano M 1995 Phys. Rev. Lett. {\bf 74}, 2812.

\noindent Konik R, Lesage F, Ludwig AWW and Saleur H 1998 cond-mat/9806334.

\noindent Kumagai K, Tsuji S, Kato M and Koike Y 1997 Phys. Rev. Lett. {\bf 78},
1992.

\noindent Kurniawan B, Tanaka H, Takatsu K, Shiramura W, Fukuda T,
Nojiri H and Motokawa M 1999 Phys. Rev. Lett. {\bf 82}, 1281.

\noindent Kuroki H and Aoki H 1994 Phys. Rev. Lett. {\bf 72} 2947.

\noindent Laukamp M, Martins GB, Gazza C, Malvezzi A, Dagotto E, Hansen P,
Lopez A and Riera J 1998 Phys. Rev. {\bf B 57}, 10755.

\noindent Lemmens P, Fischer M, Els G, G\"untherodt G, Mishchenko AS, 
Weiden M, Hauptmann R, Geibel C and Steglich F 1998 cond-mat/9810062.

\noindent Levy B 1996 Physics Today (October) page 17.

\noindent Lin HH, Balents L and Fisher MPA 1997 Phys. Rev. {\bf B 56} 6569.

\noindent Lin HH, Balents L and Fisher MPA 1998 cond-mat/9804221.

\noindent Mc Carron EM, Subramanian MA, Calabrese JC and 
Harlow RL 1988 Mater. Res. Bull. {\bf 23}, 1355.

\noindent Magishi K, Matsumoto S, Kitaoka Y, Ishida K, Asayama K, Uehara M,
Nagata T and Akimitsu J 1998 Phys. Rev. {\bf B 57}, 11533.

\noindent Martin-Delgado MA, Shankar R and Sierra G 1996 Phys. Rev. Lett. {\bf 77} 3443.

\noindent Martins G, Dagotto E and Riera J 1996
Phys. Rev. {\bf B 54}, 16032.

\noindent Martins G, Laukamp M, Riera J and
Dagotto E 1997  Phys. Rev. Lett. {\bf 78 }, 3563.

\noindent Martins G, Gazza C and Dagotto E 1999 Phys. Rev. {\bf B 59}, 13596.

\noindent Matsuda M, Katsumata K, Eisaki H, Motoyama N, Uchida S, Shapiro SM
and Shirane G 1996 Phys. Rev. {\bf B 54}, 12199.

\noindent Matsuda M, Katsumata K, Osafune T, Motoyama N, Eisaki H, Uchida S,
Yokoo T, Shapiro SM, Shirane G and Zarestky JL 1997 cond-mat/9709203,
preprint.

\noindent Matsuda M, Yoshihama T, Kakurai K and Shirane G 1998 cond-mat/9808083.

\noindent Matsuda M, Katsumata K, Eccleston R, Brehmer S 
and Mikeska H.-J. 1999 preprint.

\noindent Matsumoto S, Kitaoka Y, Ishida K, Asayama K, Hiroi Z, Kobayashi N
and Takano M 1996 Phys. Rev. {\bf B 53}, 11942.

\noindent Mayaffre H, Auban-Senzier P, Nardone M, J\'erome D, Poilblanc D, 
Bourbonnais C, Ammerahl U, Dhalenne G  and Revcolevschi A 1998 Science
{\bf 279}, 345.

\noindent Millet P, Satto C, Bonvoisin J, Normand B, Penc K, Albrecht M and
Mila F 1998 Phys. Rev. {\bf B 57}, 5005.

\noindent Mito T, Magishi K, Matsumoto S, Zheng G, Kitaoka Y, Asayama K, Motoyama
N, Eisaki H and Uchida S 1998 preprint, to appear in Physica B,
proceedings of the SCES98 conference, Paris.

\noindent Miyahara S, Troyer M, Johnston DC and Ueda K 1998 cond-mat/9807127

\noindent Miyazaki T, Troyer M, Ogata M, Ueda K and Yoshioka D 1997 J. Phys.
Soc. Jpn. {\bf 66}, 2580.

\noindent Mizokawa T, Ootomo K, Konishi T, Fujimori A, Hiroi Z, Kobayashi N
and Takano M 1997 Phys. Rev. {\bf B 55}, R13373.

\noindent Mizokawa T, Okazaki K, Fujimori A, Osafune T, Motoyama N, 
Eisaki H and Uchida S 1998 preprint.

\noindent Mizuno Y, Tohyama T and Maekawa S 1997 
J. Phys. Soc. Jpn. {\bf 66}, 937; 1997
Physica C {\bf 282-287}, 991.

\noindent Moshchalkov V, Trappeniers L and Vanacken J 1998 submitted to Europhys.
Letters. 

\noindent Motome Y, Katoh N, Furukawa N and Imada M 1996 J. Phys. Soc. Jpn. {\bf
65}, 1949.

\noindent Motoyama N, Eisaki H and Uchida S 1996 Phys. Rev. Lett. {\bf 76} 3212.

\noindent Motoyama N, Osafune T, Kakeshita T, Eisaki H and Uchida S 1997
Phys. Rev. {\bf B 55} R3386-R3389.


\noindent Nagaosa N, Furusaki A, Sigrist M and Fukuyama H 1996 J. Phys. Soc. Jpn.
{\bf 65}, 3724.

\noindent Nagata T, Uehara M, Goto J, Komiya N, Akimitsu J, Motoyama N, 
Eisaki H, Uchida S, Takahashi H, Nakanishi T and Mori N 1997 Physica C
{\bf 282-287}, 153.

\noindent Nagata T, Uehara M, Goto J, Akimitsu J, Motoyama N, Eisaki H,
Uchida S, Takahashi H, Nakanishi T and Mori N 1998a Phys. Rev. Lett. {\bf
81} 1090-1093.

\noindent Nagata T, Fujino H, Akimitsu J, Nishi M, Kakurai K, Katano S,
Hiroi M, Sera M and Kobayashi N 1998b preprint.

\noindent Nagata T, Fujino H, Ohishi K, Akimitsu J, Katano S, Nishi M and
Kakurai K 1999 preprint.

\noindent Nakamura T and Okamoto K 1998 Phys. Rev. {\bf B 58}, 2411.

\noindent Nazarenko A, Moreo A, Riera J and Dagotto E 1996 Phys. Rev. {\bf B 54} R768.

\noindent Nersesyan AA and Tsvelik AM 1997 Phys. Rev. Lett. {\bf 78} 3939.

\noindent Ng TK 1996 Phys. Rev. {\bf B 54}, 11921.

\noindent Noack RM, White SR and Scalapino DJ 1994 Phys. Rev. Lett. {\bf
73}, 882.

\noindent Normand B and Rice TM 1996 Phys. Rev. {\bf B 54}, 7180.

\noindent Normand B, Penc K, Albrecht M and Mila F 1997 Phys. Rev. {\bf B 56},
R5736.

\noindent Normand B and Rice TM 1997 Phys. Rev. {\bf B 56}, 8760.

\noindent Normand B, Agterberg DF and Rice TM 1998 cond-mat/9812211, preprint.

\noindent Ohsugi S, Tokunaga Y, Ishida K, Kitaoka Y, Azuma M, Fujishiro Y and
Takano M 1998 preprint, to appear in Phys. Rev.  B.

\noindent Ohsugi S, Magishi K, Matsumoto S, Kitaoka Y, Nagata T and
Akimitsu J 1999 Phys. Rev. Lett. {\bf 82}, 4715.

\noindent Ohta T, Onoda M, Izumi F, Isobe M, Takayama-Muromachi E 
and Hewat AW 1997 J. Phys. Soc. Jpn. {\bf 66}, 3107.

\noindent Orignac E and Giamarchi T 1998 Phys. Rev. {\bf B 57} 5812.

\noindent Osafune T, Motoyama N, Eisaki H and Uchida S 1997 Phys. Rev. 
Lett. {\bf 78}, 1980.

\noindent Osafune T, Motoyama N, Eisaki H, Uchida S and Tajima S 1999
Phys. Rev. Lett. {\bf 82}, 1313.

\noindent Owens FJ, Iqbal Z and Kirven D 1996 preprint.

\noindent Park Y, Liang S and Lee TK 1998 cond-mat/9811280.

\noindent Piekarewicz J and Shepard JR 1997 Phys. Rev. {\bf B 56}, 5366.

\noindent Poilblanc D, Tsunetsugu H and Rice TM 1994 Phys. Rev. {\bf B 50}, 6511.

\noindent Regnault LP, Boucher JP, Moudden H, Lorenzo JE, Hiess A, 
Ammerahl U, Dhalenne G and Revcolevschi A 1998 cond-mat/9809009, preprint.

\noindent Rice TM, Gopalan S and Sigrist M 1993 Europhys. Lett. {\bf 23}, 445.

\noindent Riera J and Dagotto E 1993 Phys. Rev. {\bf B 47} 15346.

\noindent Riera J 1994 Phys. Rev. {\bf B 49} 3629.

\noindent Riera J, Poilblanc D and Dagotto E 1999
 Eur. Phys. J. B {\bf 7}, 53.

\noindent Rojo A 1996 Phys. Rev. {\bf B 53} 9172.

\noindent Ruzicka B, Degiorgi L, Ammerahl U, Dhalenne G and Revcolevschi A 1998
Eur. Phys. J. {\bf B6} 301.

\noindent Sandvik AW, Dagotto E and 
Scalapino DJ 1997 Phys. Rev. {\bf B 56}, 11701.

\noindent Sato T, Yokoya T, Takahashi T, Uehara M, Nagata T, Goto J and 
Akimitsu J 1997 preprint, to be published in J. Phys. Chem. Solids.

\noindent Schmeltzer D and Bishop AR 1998 Phys. Rev. {\bf B 57} 5419.

\noindent Schulz H 1996 Phys. Rev. {\bf B 53} R2959.

\noindent Shiramura W, Takatsu K, Tanaka H, Kamishima K, Takahashi M,
Mitamura H and Goto T 1997 J. Phys. Soc. Jpn. {\bf 66}, 1900.

\noindent Siegrist T, Schneemeyer LF, Sunshine SA, Waszczak JV and  
Roth RS 1988 Mater. Res. Bull. {\bf 23}, 1429.

\noindent Sierra G, Martin-Delgado MA, Dukelsky J, White SR and Scalapino DJ 1998
Phys. Rev. {\bf B 57} 11666.

\noindent Sigrist M, Rice TM and Zhang FC 1994 Phys. Rev. {\bf B 49}, 12058.

\noindent Smolinski H, Gros C, Weber W, Peuchert U, Roth G, Weiden M and 
Geibel C 1998 Phys. Rev. Lett. {\bf 80}, 5164.

\noindent Syljuasen OF, Chakravarty S and Greven M 1997 Phys. Rev. Lett. {\bf 78} 4115.

\noindent Takahashi T, Yokoya T, Ashihara A, Akaki O, Fujisawa H, 
Chainani A, Uehara M, Nagata T, Akimitsu J and Tsunetsugu H 1997 Phys.
Rev. {\bf B 56}, 7870.

\noindent Takatsu K, Shiramura W and Tanaka H 1997 J. Phys. Soc. Jpn.
{\bf 66}, 1611.

\noindent Takenaka K, Mizuhashi K, Takagi H and Uchida S 1994 Phys. Rev. {\bf B
50}, 6534.

\noindent Takigawa M, Motoyama N, Eisaki H and Uchida S 1998 Phys. Rev. {\bf B
57}, 1124.

\noindent Tanaka H, Takatsu K, Shiramura W and Ono T 1996 J. of Phys. Soc. Jpn.
{\bf 65}, 1945.

\noindent Thurber K, Imai T, Saitoh T, Azuma M, Takano M and Chou F C 1999
cond-mat/9906141, preprint.

\noindent Troyer M, Tsunetsugu H and W\"urtz D 1994 Phys. Rev. {\bf B 50}, 13515.

\noindent Troyer M, Tsunetsugu H and Rice TM 1996 Phys. Rev. {\bf B 53}, 251.

\noindent Troyer M, Zhitomirsky M and Ueda K 1997 Phys. Rev. {\bf B 55}, R6117.

\noindent Tsuji S, Kumagai K, Kato M and Koike Y 1996 J. Phys. Soc. Jpn. 
{\bf 65}, 3474. 

\noindent Tsunetsugu H, Troyer M and Rice TM 1994 Phys. Rev. {\bf B 49}, 16078.

\noindent Tsunetsugu H, Troyer M and Rice TM 1995 Phys. Rev. {\bf B 51}, 16456.

\noindent Uchida S, Ido T, Takagi H, Arima T, Tokura Y and Tajima S 1991 Phys.
Rev. {\bf B 43}, 7942.

\noindent Uehara M, Nagata T, Akimitsu J, Takahashi H, Mori N and
Kinoshita K 1996 J. Phys. Soc. Jpn. {\bf 65}, 2764.

\noindent Weihong Z, Singh RRR and Oitmaa J 1997 Phys. Rev. {\bf B 55}, 8052.

\noindent White SR, Noack RM and Scalapino DJ 1994 Phys. Rev. Lett. {\bf 73}, 886.

\noindent Yamaji K and Shimoi Y 1994 Physica C {\bf 222} 349.

\noindent Zhai~Z, Patanjali~P, Hakim~N, Sokoloff~J, Sridhar~S, 
Ammerahl~U, Vietkine~A and

\noindent Revcolevschi A 1999 cond-mat/9903198, preprint.

\end{document}